\documentclass[aps,pre,twocolumn,showpacs,preprintnumbers,amsmath,amssymb,
superscriptaddress,floatfix,showkeys,footinbib]{revtex4-1}

\pdfoutput=1

\usepackage{graphicx}
\usepackage{dcolumn}
\usepackage{bm}
\usepackage{color} 
\usepackage{amsfonts}
\usepackage{amsmath}
\usepackage{amssymb}
\usepackage{amsthm}
\usepackage[dvipsnames]{xcolor}
\usepackage{soul}	

\DeclareMathOperator{\sgn}{sgn}

\newcommand{\unnec}[1]{}

\newcommand{\unnecc}[1]{}

\newcommand{\defp}{q} 

\begin{document} 
 
 
\title{Three-step melting of hard superdisks in two dimensions}

\author{P\'eter Gurin} 
\affiliation{Physics Department, Centre for Natural Sciences, University of 
Pannonia, P.O. Box 158, Veszpr\'em H-8201, Hungary} 

\author{Szabolcs Varga} 
\affiliation{Physics Departement, Centre for Natural Sciences, University of Pannonia, 
P.O. Box 158, Veszpr\'em H-8201, Hungary}

\author{Gerardo Odriozola} \email{godriozo@azc.uam.mx}
 \affiliation{\'Area de F\'isica de Procesos 
Irreversibles, Divisi\'on de Ciencias B\'asicas e Ingenier\'ia, Universidad 
Aut\'onoma Metropolitana-Azcapotzalco, Av. San Pablo 180, 02200 CD M\'exico, 
Mexico}

 
\begin{abstract} 
We explore the link between the melting scenarios of two-dimensional 
systems of hard disks and squares through replica-exchange 
Monte Carlo simulations of hard superdisks. The well-known melting 
scenarios are observed in the disk and square limits, while we 
observe an unusual three-step scenario for dual-shapes. We find that two 
mesophases mediate the melting: a hexatic phase and another fluid phase with a 
$D_2$ local symmetry, we call it \emph{rhombatic}, where both bond and particle 
orientational orders are quasi-long-range. Our results show that not only can 
the melting process of liquid-crystal forming molecules be complicated, 
where elongated shapes stabilize several mesophases, but also that of 
anisotropic quasispherical molecules. 
\end{abstract}

\maketitle

\section{Introduction}

Melting of two-dimensional (2D) systems with short-range (SR) interactions is 
still a matter of debate. According to simulation and experimental studies, 
three possible 2D melting scenarios have been observed so far. In the 
continuous two-step melting, there is an intermediate $x$-atic phase such as 
tetratic and hexatic~\cite{Murray87, Donev06, Han-Ha-Alsayed-Yodh_PRE_2008}. 
This transition is described by the 
Kosterlitz--Thouless--Halperin--Nelson--Young
(KTHNY) theory~\cite{Halperin-Nelson,Kosterlitz-Thouless,Young_PRB_1979}, where 
the unbinding of the topological defects is responsible for the melting. 
The second type arises when defects form strings, which leads to a 
grain-boundary induced discontinuous 
melting~\cite{Anderson2017,Karnchanaphanurach-Lin-Rice_PRE_2000}. The third 
possibility is a discontinuous $x$-atic mediated two-step melting, 
which follows one or two first-order phase 
transitions~\cite{Marcus-Rice_PRL_1996, 
Dijkstra_SOFTM_2014,Thorneywork17}. The first scenario 
corresponds to the melting of hard squares with four-fold symmetry (the 
intermediate phase is tetratic)~\cite{Wojciechowski04}, the second one appears 
for hard pentagons~\cite{Schilling05}, and the third one arises in the system 
of hard disks (the intermediate phase is hexatic)~\cite{Bernard_PRL_2011, 
Engel_PRE_2013}.

To find a bridge between different melting scenarios, it is useful
to study the phase behavior of model potentials, where the softness
of the interaction or the shape of the particle can be continuously varied. For 
instance, a family of soft disk models can be constructed with repulsive 
power-law functions, which shows a weakening of the first-order 
isotropic--hexatic phase transition with the softening of the 
interaction. The weakening process ends by turning the discontinuous two-step 
melting into the continuous two-step KTHNY scenario~\cite{Kapfer2015}. Also, 
adding attractive interactions to the hard ones completely destabilizes the 
hexatic phase and changes the nature of the discontinuous two-step melting into 
a one-step transition~\cite{Li-Ciamarra_PRL_2020}. Finally, all three melting 
scenarios are observed for hard regular polygons~\cite{Anderson2017}. However, 
the particle shape cannot be changed continuously in this family of models, and 
so, polygons are not suitable to study the competition of conflicting $x$-atic 
phases. For this purpose, we smoothly deform the disks into squares. For 
dual-shapes, the system resolves this imposed conflict by producing a complex 
melting process.

\section{Methods}

We study the change between the 
discontinuous and continuous 
two-step melting scenarios and the occurrence of solid-to-solid transitions 
in a 2D system of hard superdisks. A superdisk is defined by 
$|2x/\sigma|^\defp+|2y/\sigma|^\defp \leq 1$, where $\defp$ is the deformation 
parameter and $\sigma$ is the side length of the circumscribing square. See
Fig.~S1 of the Supplemental Material (SM) section placed at the end of this 
text. We perform replica-exchange Monte 
Carlo simulations~\cite{Lyubartsev92,hukushima96,Okabe01,Gerardo_PCCP_2018} for 
$2 \leq \defp \leq 20$, where $\defp=2$ corresponds to the hard disk limit, and 
$\defp=20$ practically yields squares. We generally use $N_r=100$ replicas and 
$N=196$ particles, and additionally, in the most interesting cases, $N_r=40$ 
and $N=6400$. The details of the Monte Carlo simulations and the algorithm 
employed to avoid overlaps between superdisks are given at the SM. 
Here we show that the system of hard superdisks can melt in three continuous 
steps, where two $x$-atic phases, namely rhombatic and hexatic, mediate between 
the solid and isotropic liquid phases. This new scenario occurs for 
dual-shapes, $5\lesssim\defp\lesssim7$, which are halfway between the 
disk and the square.

\section{Results}

\begin{figure*}[t!]
\centering
\includegraphics[width=0.98\linewidth]{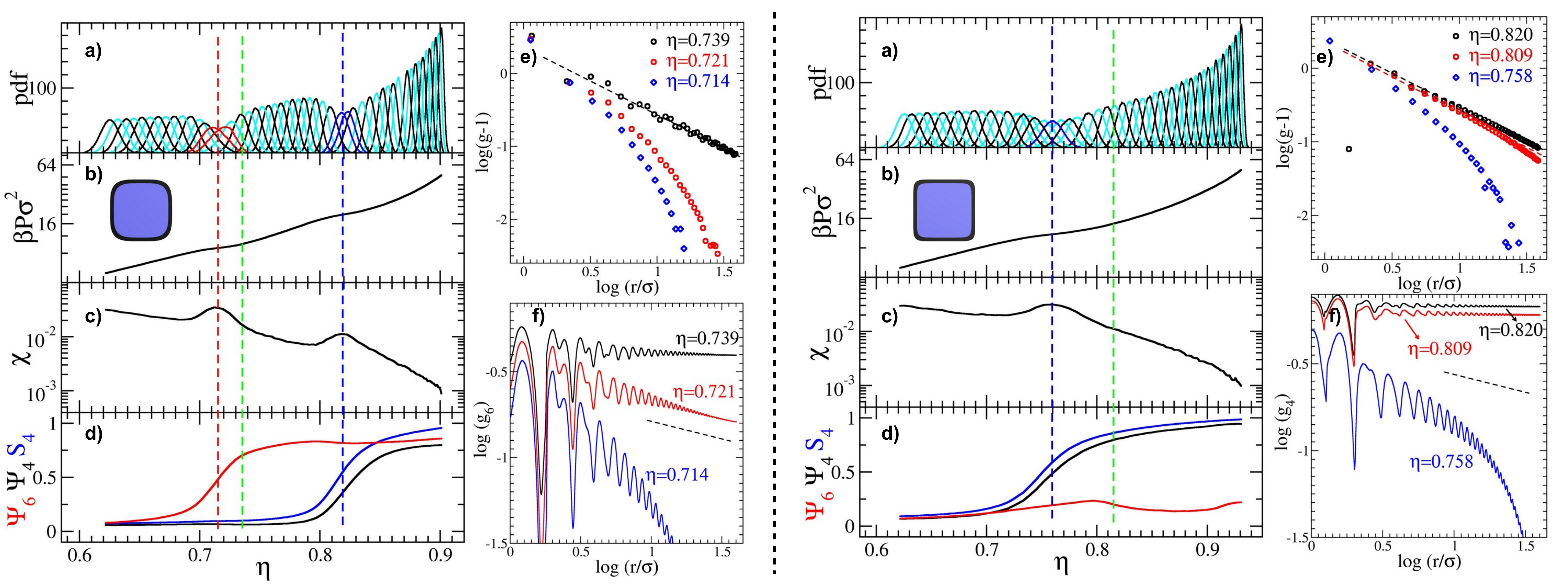}
\caption{Melting of hard superdisks with $\defp=4$ (left panel) and $\defp=8$ 
(right panel). a) Probability density functions, where red and blue lines are 
employed to highlight the histograms close to the isotropic--hexatic and the 
plastic solid--rhombic solid transitions, respectively. b) Dimensionless 
pressure, $\beta P \sigma^2$, c) dimensionless isothermal compressibility, 
$\chi$, and d) global order parameters, $\Psi_6$ (red line), $\Psi_4$ 
(black line), and $S_4$ (blue line), as a function of the packing fraction, 
$\eta$, for systems of $N=196$ superdisks. The vertical dashed red and blue 
lines signal the development of hexatic and rhombic structures (QLR six-fold 
and four-fold bond-orientational order), respectively, and the green line 
signals the development of QLR positional order. Panels e) show the peaks of 
$\log(g-1)$ and panels f) show $\log(g_6)$ (left) and $\log(g_4)$ (right) as a 
function of $\log(r/\sigma)$ for a system of $N=6400$ superdisks. The black 
dashed lines in panels f) have a slope of $-1/4$. The corresponding superdisk's 
shapes are shown as insets.}
\label{fig:case4&8}
\end{figure*}

\begin{figure*}[t!]
\centering
\includegraphics[width=0.98\linewidth]{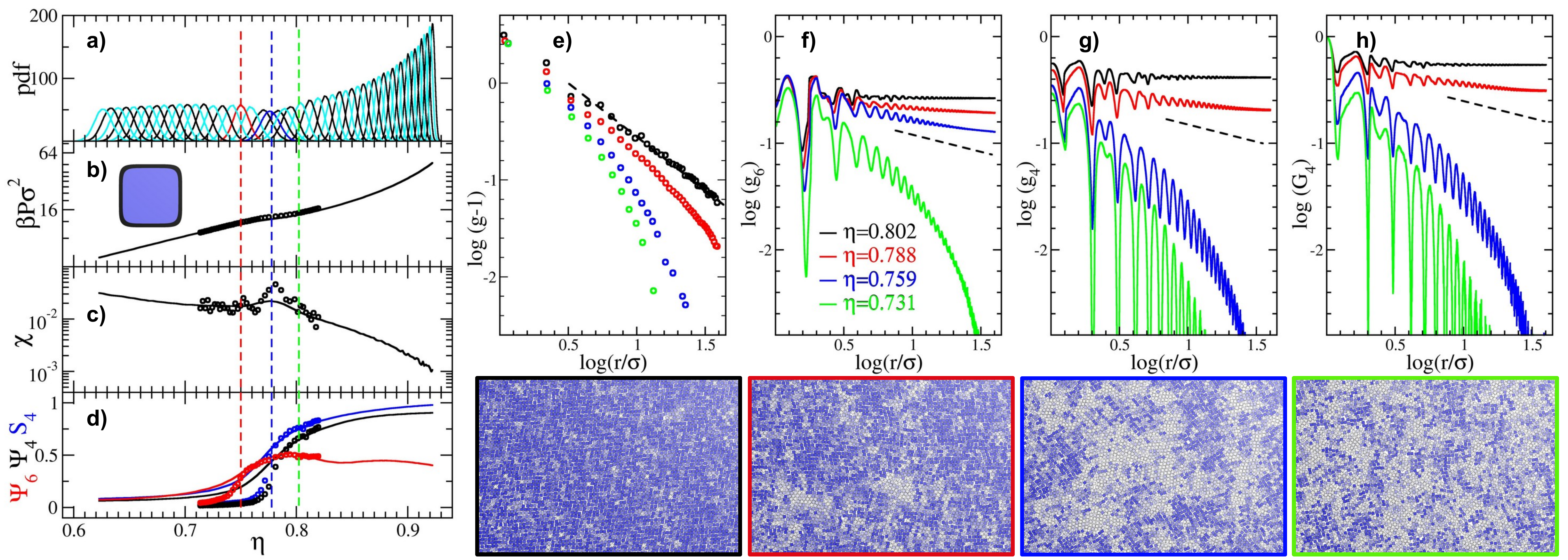}
\caption{Melting of superdisk with $q=6$. See the caption of 
Fig.~\ref{fig:case4&8} for the meaning of the a)-d) panels.
We also include the $N=6400$ data as symbols in panels b)-d). Panels
e)-h) show the different correlation functions defined in the text. The color 
of the curves in panels e)-h) matches the color of the frames of the snapshots. 
Particles painted blue and white are oriented parallel and forming an angle of 
$\pi/4$ with the snapshot director, respectively. Intermediate cases are 
painted with intermediate tones. }
\label{fig:case6.0}
\end{figure*}

We present our results for $\defp=4$ and 8 in
Fig.~\ref{fig:case4&8}, which have some common features with the
melting scenarios of the corresponding hard disk ($\defp=2$) and square 
limits ($\defp\rightarrow\infty$)~\cite{Bernard_PRL_2011,Wojciechowski04}.
In Fig.~\ref{fig:case6.0} we show the results for
$\defp=6$ demonstrating the new three-step melting scenario. Also, the case of 
$\defp=2.5$ is depicted in Fig.~S5. In all of these figures, 
panels a) show the probability density functions (PDFs), which are distorted 
from the Gaussian-shape in the vicinity of a first-order transition. For 
instance, this is the case of the isotropic--hexatic 
transition occurring for low $\defp$ values (see the $\defp=2.5$ case in 
Fig.~S5). We have found that the distortion from the 
Gaussian-shape of the PDFs weaken up to $\defp \approx 4$. Indeed, panel a) 
on the left-hand side of Fig.~\ref{fig:case4&8} still shows slight deviations 
from the Gaussian-shape for those PDFs close to the isotropic-hexatic 
transition. For this reason, we estimate this first-order transition to end 
slightly above $\defp=4$. This is consistent 
with the equations of state shown in panels b), given that they have a plateau 
for $\defp\lesssim4$, 
and the pressure is a strictly monotonic function of the density for 
$\defp > 4$. In panels c), we present the dimensionless isothermal 
compressibility, $\chi= d \rho/d (\beta P)$. The peak 
of $\chi$ survives for $\defp>4$ and locates the continuous 
isotropic--hexatic transition (see Fig.~S6), which is the 
general behavior of a continuous symmetry-breaking transition. The maximum of 
$\chi$ determines not only the location of the isotropic--hexatic transition, 
signaled by the vertical red dashed lines crossing from panels a) to d) of all 
these figures, but also the location of other continuous transitions (the blue 
dashed lines). We show the $\defp$ dependence of the $\chi$ 
curves in Fig.~S6, where one can see maxima producing a 
Y-shaped pattern that is important for the construction of the global phase 
diagram. 

Panels d) of Figs.~\ref{fig:case4&8}, \ref{fig:case6.0}, and S5, show the 
global tetratic order parameter, $S_4=\langle 
\frac{1}{N}|\int S_4(r) dr|\rangle$, and the global $n$-fold bond-orientational 
order parameters, $\Psi_n=\langle \frac{1}{N}|\int \psi_n(r) dr|\rangle$ with 
$n=4$ and 6. Here, $S_4(r)=\sum_j e^{i4\alpha_{j}}\delta(r-r_j)$ is the 
local particle orientational order parameter, 
$\psi_n(r)=\sum_j\frac{1}{n}\sum_{k=1}^n e^{in\phi_{jk}}\delta(r-r_j)$ is 
the local bond-orientational order parameter, $r_j$ and $\alpha_j$ describe the 
position and orientation of particle $j$, and $\phi_{jk}$ is the angle of the 
bond linking particle $j$ and its neighbor $k$. We can see the growth of these 
functions at the transitions. To describe the phase behavior more precisely, in 
panels e) we show the positional order of the particles through the peaks of the 
radial distribution function, $g(r)$. The height of the peaks keeps constant 
with 
increasing distance only in the infinite pressure limit and decays algebraically 
for a solid phase, which is referred to as quasi-long-range (QLR) behavior. 
In the case of a liquid, this function exhibits SR correlations (exponential 
decay). Furthermore, in panels f), we show the $n$-fold 
bond-orientational correlation function, $g_n(r)=\langle \psi^{*}_n(r) 
\psi_n(0)\rangle$, where $n=4$ and 6. These functions show a constant behavior 
at large distances in a solid phase, decay algebraically in a bond-ordered 
$x$-atic fluid, and decay exponentially in the isotropic phase. Finally, we show 
in Fig.~\ref{fig:case6.0} h) the correlation function of the 
particle orientations, $G_4(r)=\langle S_4^{*}(r) S_4(0) \rangle$. 
According to the KTHNY theory~\cite{Kosterlitz-Thouless, Halperin-Nelson}, 
particle orientational functions decaying faster than $\sim r^{-1/4}$ are SR. 
Due to this reason, we include straight lines with this slope in panels f) of 
Fig.~\ref{fig:case4&8} and in panels f), g), and h) of Fig.~\ref{fig:case6.0}. 
The $x$-atic--solid continuous transitions, which are indicated by vertical 
green dashed lines crossing from panels a) to d), are obtained from the 
long-distance behavior of $g$, $g_4$, and $g_6$. Note that $N=6400$ 
particles may not be enough to access the quasi-long-range behavior of the 
system. This may lead to a slight underestimation of this transition density.

In cases $4 \lesssim \defp \lesssim 5$ and $\defp \gtrsim 7$, the 
melting follows the continuous two-step scenario (see Fig.~\ref{fig:case4&8}). 
In the first step, the QLR positional order is destroyed and the 
long-range bond-orientational order becomes QLR. At this point, the solid 
turns into an $x$-atic fluid phase. In the second step, every global order 
parameter yields low values, while all correlation functions become SR. Here, 
the $x$-atic turns into an isotropic phase, which is also signaled by a 
compressibility peak. The difference between the $4 \lesssim \defp \lesssim 5$ 
and $\defp \gtrsim 7$ cases stems from the symmetries of the solid phases, which 
are $D_6$ and $D_2$ ($D_n$ denotes the dihedral group), respectively. The 
primitive 
unit cell of the later lattice is a rhombus, and so we call it 
rhombic solid (see the SM). As the intermediate mesophases locally 
inherit the properties of the solid phases, there appears a \emph{rhombatic} 
phase for $\defp \gtrsim 7$ instead of the hexatic phase for $4 \lesssim \defp 
\lesssim 5$. An important difference between hexatic and rhombatic phases is 
that the orientations of the particles show QLR order in the latter case ($S_4 
> 
0$ and $G_4$ decays algebraically), while it is disordered in the former case 
($S_4 \approx 0$ and $G_4$ decays exponentially). Moreover, both the 
four-fold and the six-fold bond-orientation show QLR order in the rhombatic 
phase ($g_4$ and $g_6$ decay algebraically), while only the six-fold 
bond-orientation exhibits QLR order in the hexatic phase. Note that the rhombic 
and rhombatic phases yield the square and tetratic phases in the 
$\defp\rightarrow\infty$ limit, respectively. Therefore, 
the melting processes of hard squares and for $\defp=8$ are qualitatively the 
same, where the only difference is the replacement of 
the $D_4$ symmetry of the square lattice with the $D_2$ symmetry of the more 
general rhombic structure. It is worth mentioning that the rhombatic phase is 
not observed in other families of rounded squares, where the tetratic or the 
hexatic phases mediate the melting~\cite{Avendano12,Zhaglin_ChinPhysB_2018}. 
This may be because the superdisk does not have parallel straight sides 
contrasting with the 
other models.

We should emphasize the deviation of the $4 \lesssim \defp \lesssim 5$ cases 
from the disk limit. The isotropic--hexatic transition is continuous for
$4 \lesssim \defp \lesssim 5$ and discontinuous for hard
disks~\cite{Bernard_PRL_2011}. This fact closely resembles the case of soft 
disks~\cite{Kapfer2015}. Superdisks can get closer than the 
diagonal of the particle while being orientationally unfrozen, which renders the
particles to behave as soft disks. Indeed, the continuous two-step melting 
scenario can also be observed in the system of soft disks with $V(r)\sim r^{-n}$
potential for $n<6$~\cite{Kapfer2015}. Thus, increasing $\defp$ 
effectively changes the range of the contact distance between two superdisks 
to transform the isotropic--hexatic transition from discontinuous to continuous.

For $\defp \lesssim 5$, the symmetry of the solid phase can be reduced with 
increasing density. It manifests with an additional peak of $\chi$ and the 
growth of LR particle orientational order. This solid--solid transition is 
continuous and occurs between a plastic (or rotator) solid with a hexagonal 
structure and the rhombic structure. Note that this transition is confirmed 
experimentally in the monolayer of colloidal rounded squares~\cite{Zhao2011}. 
During this transition the orientational entropy stabilizes the low-density 
plastic phase, while the packing entropy prevails over the orientational one in 
the high-density rhombic phase. 

The stabilization of the hexatic and the rhombatic mesophases between the 
isotropic fluid and the rhombic solid phases appears for shapes halfway between 
the disk and the square (e.g. $\defp=6$). Here, the rhombic crystal 
melts into the rhombatic fluid, then the 
rhombatic mesophase transforms into the hexatic fluid, and finally, the hexatic 
fluid turns isotropic. Fig.~\ref{fig:case6.0} shows the details of this 
continuous three-step scenario. Panels e)--h) 
confirm the expected behavior of the correlation functions in the rhombic solid 
(black), rhombatic (red), hexatic (blue), and isotropic (green) phases, 
respectively. Panel d) proves that these behaviors are consistent with 
the values of the order parameters. Comparing this case with the continuous 
two-step melting, it turns out that the additional third step is the 
rhombatic--hexatic intermediate transition, where $G_4$ and $g_4$ become 
SR simultaneously, but the six-fold correlations remain QLR.

As found for the other cases, the symmetry-breaking transitions are accompanied 
by $\chi$ peaks (isotropic--hexatic and hexatic--rhombatic), while the
rhombic--rhombatic transition can be detected only from the
analysis of the correlation functions. Fig.~\ref{fig:case6.0}c) also shows
that the highest peak of $\chi$ corresponds to the hexatic--rhombatic
transition, where not only the local symmetry-breaking occurs, but
the topological defect structure transforms from disclinations
and dislocations into localized point defects~\cite{Anderson2017}. The 
snapshots of Fig.~\ref{fig:case6.0} highlight the differences in the structures 
of these phases.

Dual-shape superdisks are not anisotropic enough to
induce a direct isotropic--rhombatic transition. Interestingly, before the 
positional freezing, an additional transition happens between two different 
bond ordered fluids. As the orientational forces (packing entropy) become more
and more dominant, they give rise to the growth of the orientational order 
without the development of a QLR positional order, yielding the rhombatic 
fluid. This process is similar to the destabilization of the plastic solid 
in favor of the rhombic solid but without the existence of QLR positional order. 
At even higher densities, the rhombic solid evolves from the rhombatic, 
analogously to the tetratic-square solid transition for $\defp \rightarrow 
\infty$. The complete scenario is similar to the two-step melting predicted by 
the KTHNY theory, in the sense that all three steps are continuous and that the 
topological defects mediate the transitions.

\begin{figure}[t!]
\centering
\includegraphics[width=0.98\linewidth]{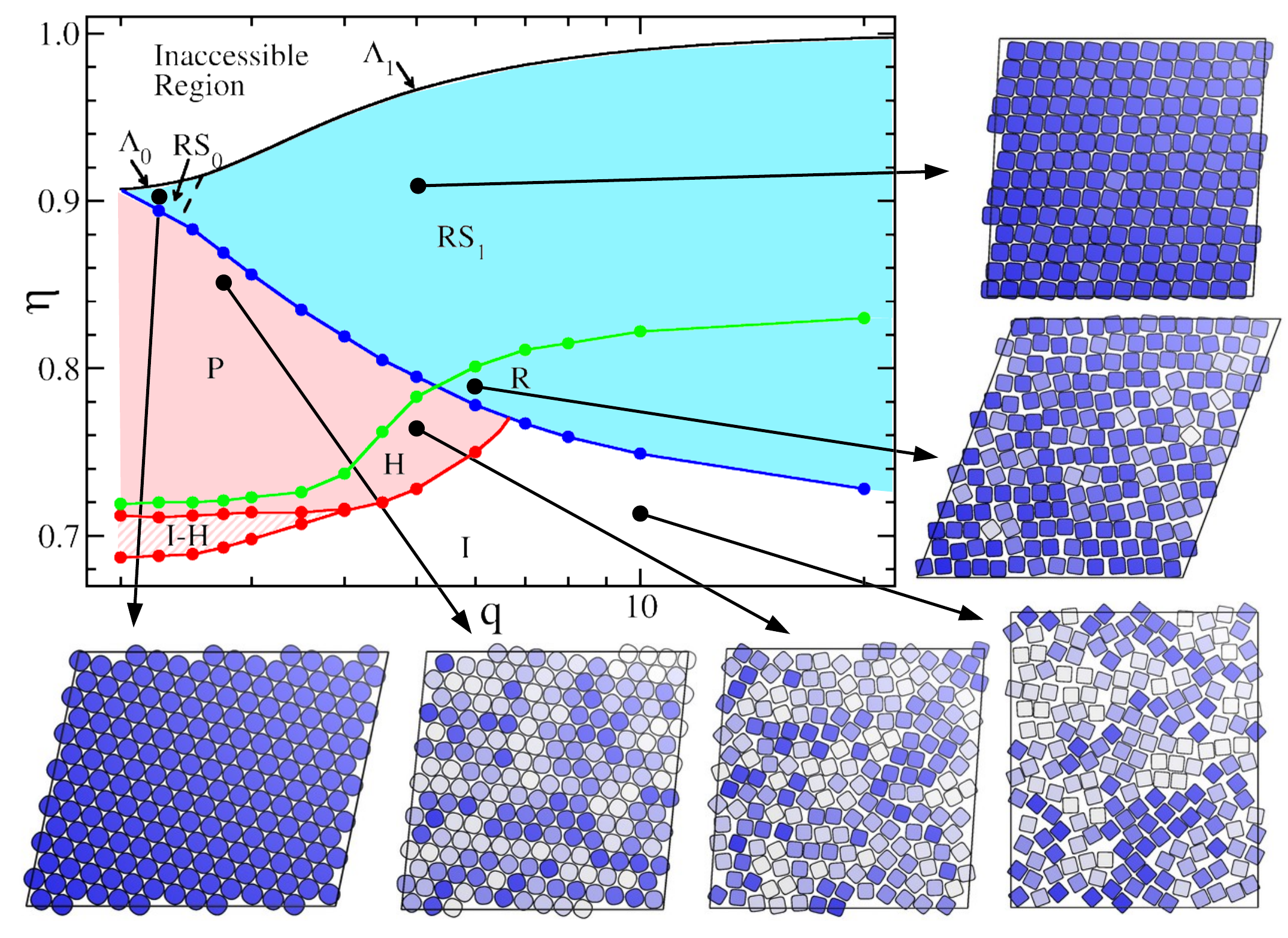}
\caption{Phase diagram of superdisks in the packing
fraction--deformation parameter ($\eta$ vs. $\defp$) plane. The
meaning of the labels is I: isotropic fluid, H: hexatic fluid, R:
rhombatic fluid, I--H: coexistence region, P: plastic solid (or
hexagonal rotator), and RS$_0$ and RS$_1$: rhombic solids. These latter
phases produce the $\Lambda_0$ and $\Lambda_1$ rhombic crystals~\cite{Jiao08} 
in the infinite pressure limit. The colors of the phase boundaries correspond 
to the colors of the vertical dashed lines in panels
a)--d) of Fig.~\ref{fig:case4&8}, \ref{fig:case6.0}, and S5. The white, light 
red, and light blue backgrounds denote the phases with full rotational, local 
$D_6$, and local $D_2$ symmetries, respectively. }
\label{fig:phase-diagram}
\end{figure}

Fig.~\ref{fig:phase-diagram} gathers and 
summarizes the information from all studied $\defp$ values as a phase diagram. 
It is convenient to start with the close-packing structures of the hard 
superdisks. Two close-packing lattice arrangements are conjectured as optimal 
$\defp\geq2$: $\Lambda_0$ and $\Lambda_1$~\cite{Jiao08}. Both of them are 
centered rectangular Bravais lattices, which can also be described by rhombic 
primitive unit cells (see the SM). $\Lambda_0$ and $\Lambda_1$ have different
lattice parameters, and give the optimal packings below and
above $\defp\approx 2.572$, respectively. As our simulation results
reproduce these close-packing structures in the high-pressure
limit (see the SM), we call these phases rhombic solid 0
and 1 (RS$_0$ and RS$_1$). The lattice angle, $\theta$, goes to
$\pi/3$ when $\defp\rightarrow2$, thus the close-packing structure
is hexagonal in the disk limit. Then, $\theta$ slowly varies with 
increasing $\defp$ up to yield a discontinuity at $\defp\approx 2.572$, when 
$\Lambda_1$ replaces $\Lambda_0$, and then increases monotonically up to
$\pi/2$ as $\defp\rightarrow\infty$. Between the RS$_0$ and the 
RS$_1$, there is a transition indicated by a dashed black line 
in Fig.~\ref{fig:phase-diagram} (for more details 
see the SM).

The stability region of the solid 
depends weakly on $\defp$, which is due to the fact that both the 
fluid-solid transition curve and the maximal packing fraction one have a similar 
shape. That is, the curvatures of both curves change from convex to concave 
with increasing $\defp$, and the distance between them is more or less the same 
for all $\defp$. However, the packing fraction window of the hexatic 
phase widens with increasing $\defp$ for $4\lesssim\defp\lesssim5$, while 
the opposite occurs to the rhombatic phase with decreasing $\defp$. Recall that 
the rhombatic phase inherits the local structure of the RS$_1$, and hence, it 
also smoothly transforms into a tetratic phase with increasing $\defp$. The 
phase diagram contains both the hexatic and the rhombatic phases in a 
relatively narrow window, $5\lesssim\defp\lesssim7$. Here, the hexatic phase 
always occurs at lower densities than the rhombatic fluid because the 
local symmetry of the emerging phase is reduced with increasing density, as the 
$D_2$ is a subgroup of the $D_6$. Note that in this $\defp$ region the system 
is capable of producing a local $D_6$ symmetry with a relatively large value of 
$\defp$, which does not occur for $\defp \gtrsim 7$. Indeed, the local 
$D_2$ symmetry is favored with increasing $\defp$ and density, contrasting with 
the $D_6$, which appears at intermediate densities and for low $\defp$ 
values. The $D_2$ region includes the RS$_0$, RS$_1$, and R, 
whereas the $D_6$ contains the H, the P, and partially the 
I-H coexistence.  

\section{Conclusions}

We found that the transition between the discontinuous
and continuous melting scenarios is not smooth when the symmetry of
the particle is changed from circular to fourfold. P transforms into I passing 
through H for weak deformations ($\defp\lesssim5$), while 
RS$_1$ melts into R before reaching I for square-like shapes 
($\defp\gtrsim7$). With increasing $\defp$, the first-order I--H 
transition weakens and becomes KTHNY-type continuous for $\defp\gtrsim4$ as the 
system of freely rotating superdisks behaves similarly to that of soft disks
~\cite{Kapfer2015}. The interplay between the H and the R
mesophases manifest in the region $5\lesssim\defp\lesssim7$, which does not
produce a first-order transition but an additional R--H
continuous step entering in-between the two-step melting
scenario. This behavior is very different from that of the regular 
polygons, where a single-step process links the continuous and 
discontinuous melting scenarios with varying the number of
vertices~\cite{Anderson2017}. In light of previous
simulations~\cite{Avendano12,Anderson2017,Zhaglin_ChinPhysB_2018},
we believe that the fourfold particle shape without parallel sides
is responsible for the formation of the rhombatic phase, which in turn leads to 
the three-step melting processes. For liquid crystalline forming molecules, 
where the elongated and flat shapes are responsible for the formation of more 
than one mesophases, the melting process can have more than two steps. Our 
results prove that the curvature of the particle can be as important as the 
aspect ratio to induce a many-step melting process. We hope that these findings 
can be tested experimentally in colloidal silica superball systems. This type of 
system was already employed to access the solid-solid transitions appearing for 
$\defp<4$~\cite{Zhao2011,Meijer-et.al_NAT.COMM_2017,Rossi5286}. Indeed, 
monolayers of these particles confirmed the spontaneous formation of 
$\Lambda_0$ and $\Lambda_1$-like lattices~\cite{Meijer-et.al_LANGMUIR_2019}. 
To test our findings, the range of the deformation parameter should be 
kept around six.

\renewcommand\thefigure{S\arabic{figure}}
\renewcommand\thetable{S\arabic{table}}
\renewcommand\theequation{S\arabic{equation}}
\setcounter{figure}{0}

\begin{figure}[t!]
\centering
\includegraphics[width=0.98\linewidth]{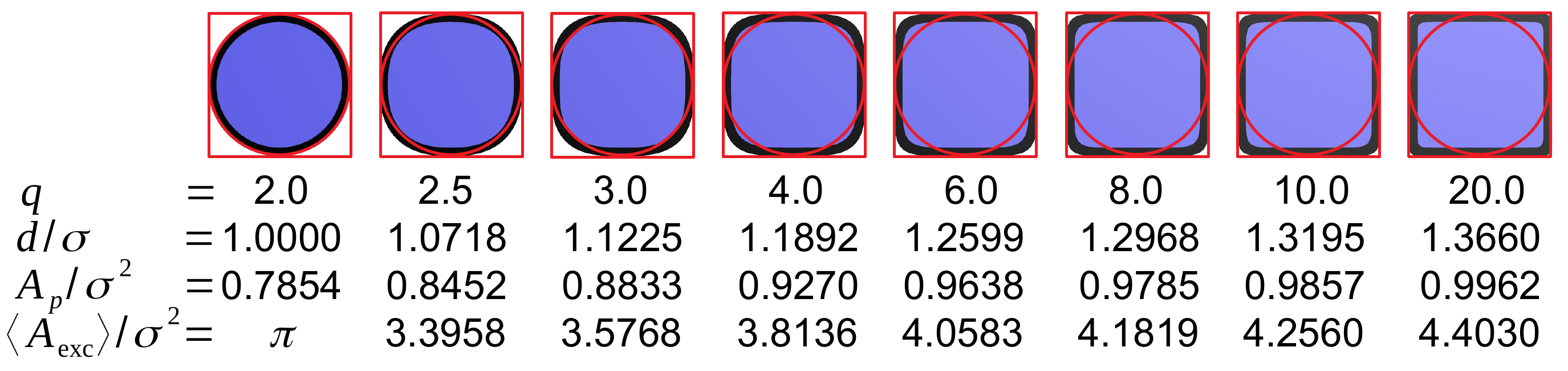}
\caption{The deformation parameter $\defp$ dependence of the superdisk' 
shape, diagonal length, area, and excluded area. The circumscribing square and 
the inscribing disk are shown in red for comparison.   }
\label{fig:q-shape}
\end{figure}


\section{Aknowledgements}

We thank the finalcial support of several sources: CONACyT through project 
A1-S-9197, the National Research, Development, and Innovation Office -- K 
124353 (GP and VS), and Fundaci\'on Marcos Moshinsky (GO).  

\section{Supplemental Material}

In this supplemental section, we first give some details on the 
superdisk shape, and then we introduce the contact algorithm 
employed to effectively detect the distance of the closest approach between two
equal superdisks. Following, we provide some details on the
simulations and present results obtained for very high pressure
to yield a twofold purpose. On the one hand, we can check the
correctness of our implementation. On the other hand, we can confirm
our algorithm leads to the conjectured optimal structures given in
[PRL, 100, 245504 (2008)]. Finally, we show details of the outcomes
for cases with $\defp=2.5$, and our all simulation results for
the isothermal compressibility and the global order parameters.

\subsection{Superdisk's shape}

\begin{figure}[t!]
\centering
\includegraphics[width=0.8\linewidth]{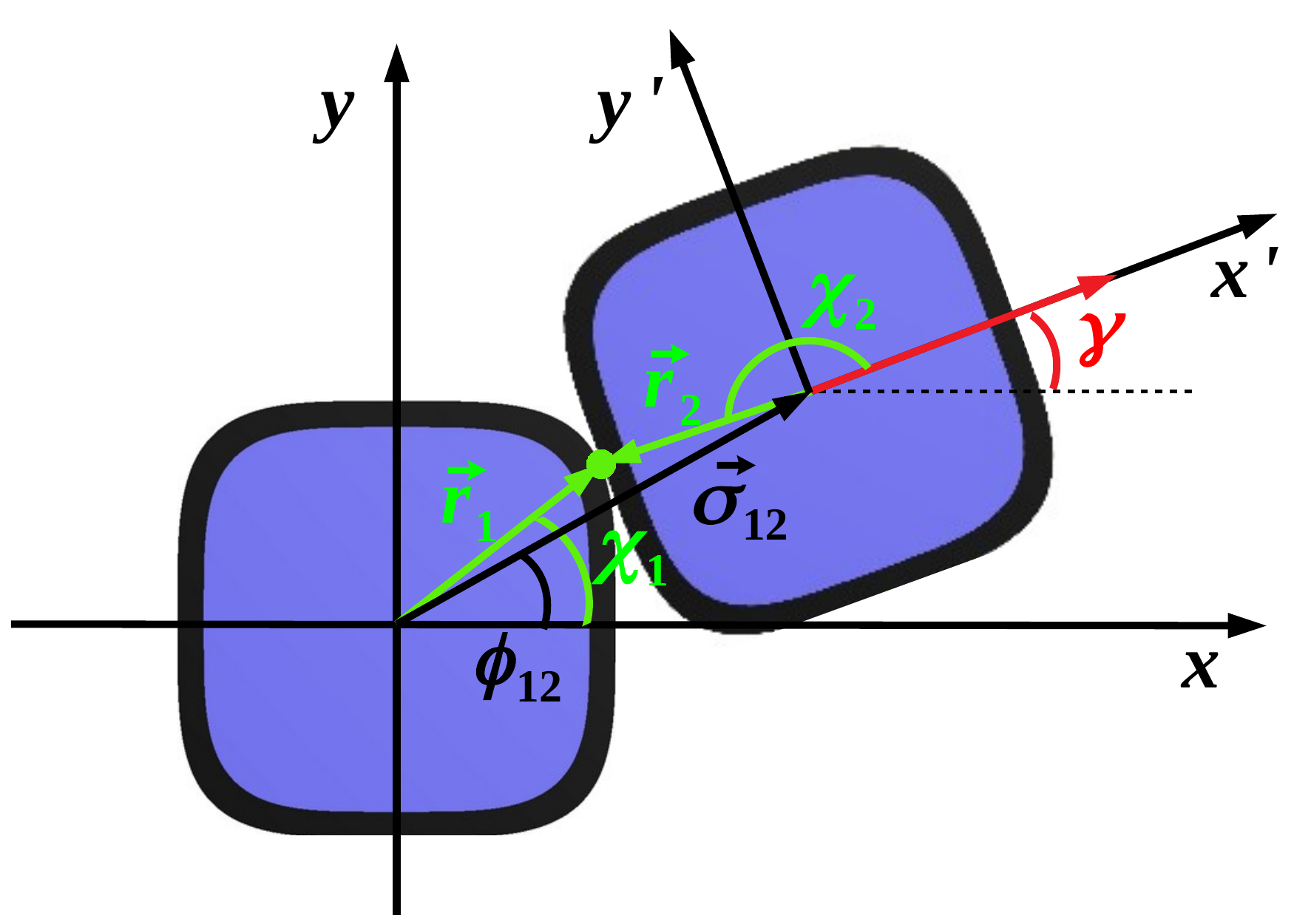}
\caption{Scheme of two superdisks in contact. The $(x,y)$ and $(x',y')$ 
reference frames lay on top of the centers of paricles 1 and 2, respectively. 
The axes of these frames coincide with the principal axes of the corresponding 
paricles. $\vec{\sigma}_{12}$ is the vector joining the centers of the 
particles, $\phi_{12}$ is the angle between $\vec{\sigma}_{12}$ and the 
$x$-axis, i.e, the bond orientation, $\vec{r}_i$ is the vector joining the 
center of particle $i$ with the contact point, $\chi_i$ is the angle between 
the vector defining the direction of particle $i$ and $\vec{r}_i$, and $\gamma$ 
is the relative orientation of particle 2 respect to particle 1.}
\label{fig:contact-superdisks}
\end{figure}

As mentioned in the main text, superdisks are defined by the set of $(x, 
y)$ points fulfilling the expression, $|2x/\sigma|^\defp+|2y/\sigma|^\defp \leq 
1$, $\sigma$ being the side length of the circumscribing square. The 
deformation parameter, $\defp$, shapes the particle as shown in 
Fig.~\ref{fig:q-shape} 
from a disk, $\defp = 2$, to a square, $\defp \rightarrow \infty$, passing 
through intermediate forms. This figure depicts how the superdisk’s diagonal, 
$d = \sigma \sqrt{ 2^{1 - 2/\defp}}$, its area, $A_p= (\sigma \Gamma(1 + 
1/\defp))^2 /\Gamma(1 + 2/\defp)$, and its average excluded area, $\langle 
A_{\rm{exc}}\rangle$, monotonically increases with $\defp$.

\subsection{Contact algorithm}

We have implemented the effective superdisk--superdisk distance of the
closest approach based on the calculation of the average excluded area
explained elsewhere~\cite{Mizani20}. Here we simplify the equations to
deal with superdisks instead of superellipses. For this purpose, we
define the vectors and distances shown in
Fig.~\ref{fig:contact-superdisks}. We also define the following
functions, which can be evaluated for a given $\gamma$ and $\chi_2$
\begin{widetext}
\begin{eqnarray} \label{eq:1}
K^2(\chi_2) &=& 
|\cos(\chi_2)|^{2(\defp-1)}+(\sigma/2)^{2\defp}|\sin(\chi_2)|^{2(\defp-1)} ,
\nonumber \\
F(\gamma,\chi_2)&=&-\cos(\gamma)\cos(\chi_2)|\cos(\chi_2)|^{\defp-2}
+(\sigma/2)^\defp 
\sin(\gamma)\sin(\chi_2)|\sin(\chi_2)|^{\defp-2} , \nonumber \\
G(\gamma,\chi_2)&=& -\sin(\gamma)\cos(\chi_2)|\cos(\chi_2)|^{\defp-2} 
-(\sigma/2)^\defp 
\cos(\gamma)\sin(\chi_2)|\sin(\chi_2)|^{\defp-2} . 
\end{eqnarray}
\end{widetext}
In turn, these quantities are employed to yield
\begin{equation} \label{eq:2a}
a(\gamma,\chi_2) = \left [ (\sigma/2)^{2\defp} 
\frac{(F(\gamma,\chi_2))^2}{K^2(\chi_2)-(F(\gamma,\chi_2))^2} \right 
]^{\frac{1}{\defp-1}}
\end{equation}
and
\begin{equation} \label{eq:2b}
b(\gamma,\chi_2) = \left [ (\sigma/2)^{-2\defp} 
\frac{(G(\gamma,\chi_2))^2}{K^2(\chi_2)-(G(\gamma,\chi_2))^2} \right 
]^{\frac{1}{\defp-1}}. 
\end{equation}
Then, we can get $\cos(\chi_1)$ and $\sin(\chi_1)$ from
\begin{equation} 
\cos(\chi_1) = 
\mathrm{sgn}(F(\gamma,\chi_2))\sqrt{\frac{a(\gamma,\chi_2)}{1+a(\gamma,\chi_2)}}
,
\label{cos_chi_1}
\end{equation}
and
\begin{equation}
\sin(\chi_1) = 
\mathrm{sgn}(G(\gamma,\chi_2))\sqrt{\frac{b(\gamma,\chi_2)}{1+b(\gamma,\chi_2)}}
.
\label{sin_chi_1}
\end{equation}
Once knowing the cosines and sines of angles $\chi_k$, with $k=1$ or $2$, we 
can finally obtain 
\begin{equation} \label{eq:4}
r_k(\chi_k)= \left [ \left ( \frac{|\cos(\chi_k)|}{\sigma/2} \right )^\defp 
+\left ( \frac{|\sin(\chi_k)|}{\sigma/2} \right )^\defp \right 
]^{-\frac{1}{\defp}}.  
\end{equation}
Given that we are working on the reference frame set on top of particle 1, we 
get
\begin{align} \label{eq:5}
\vec{\sigma}_{12} = [ r_1\cos(\chi_1)-r_2\cos(\chi_2+\gamma) ] 
\vec{e}_x \nonumber \\ + [r_1\sin(\chi_1)-r_2\sin(\chi_2+\gamma) ] 
\vec{e}_y . 
\end{align}

Therefore, via Eqs.~(\ref{cos_chi_1}-\ref{eq:5}) one can express
$\sigma_{12}=|\vec{\sigma}_{12}|$ and $\phi_{12}=\arctan{\left
  (\frac{\sigma_{12y}}{\sigma_{12x}}\right )}$ as a function of
$\gamma$ and $\chi_2$. However, it should be noted that $\gamma$ and
$\phi_{12}$ are directly defined by the position and orientation of
both particles, but not $\chi_2$. Thus, for given values of $\gamma$
and $\phi_{12}$, one needs to vary $\chi_2$ to get the desired value
of $\phi_{12}$. This is achieved by an iterative procedure until
convergence.

Alternatively, one can also start an iterative procedure with $\gamma$
and $\chi_1$ to find the value of $\chi_1$ that corresponds to
$\phi_{12}$. For this purpose, we need expressions for $\cos(\chi_2)$
and $\sin(\chi_2)$ as a function of $\chi_1$. We can easily get these
expressions by rotating an angle $\gamma$ the second particle respect to the 
first one (the first particle is rotated $-\gamma$ respect to the second) and 
by fixing the reference frame on top of particle 2 instead of particle 1. We 
get
\begin{align} \label{eq:cossin}
\cos(\chi_2) = 
\sgn(F(-\gamma,\chi_1))\sqrt{\frac{c(-\gamma,\chi_1)}{1+c(-\gamma,\chi_1)}}
\, , \nonumber  \\
\sin(\chi_2) = 
\sgn(G(-\gamma,\chi_1))\sqrt{\frac{b(-\gamma,\chi_1)}{1+b(-\gamma,\chi_1)}}
 \, ,
\end{align}
and finally, 
\begin{align} \label{eq:6}
\vec{\sigma}_{12} = [ r_1\cos(\chi_1-\gamma)-r_2\cos(\chi_2) ] \vec{e}_{x'}
\nonumber \nonumber \\ +   [ r_1\sin(\chi_1-\gamma)-r_2\sin(\chi_2) ] 
\vec{e}_{y'}. 
\end{align}
Thus, we can express
$\sigma_{12}=|\vec{\sigma}_{12}|$ and $\phi_{12}=\gamma+\arctan{\left
  (\frac{\sigma_{12y'}}{\sigma_{12x'}}\right )}$ as a function of
$\gamma$ and $\chi_1$.
Note that when $\phi_{12}(\chi_2,\gamma)$ strongly varies with $\chi_2$,
$\phi_{12}(\chi_1,\gamma)$ varies smoothly with $\chi_1$. Thus, we are using
both routes, which is convenient to avoid numerical issues.

We tabulate $\sigma_{12}(\phi_{12},\gamma)$ to avoid the iteration
procedure through the simulations. We are setting a step increase for
both angles of 0.005 rad and performing a linear interpolation of the
tabulated values. Errors of $\sigma_{12}$ are always lower than
$0.1\%$ for all $\defp$ values here studied ($\defp=20.0$ produces the
largest deviations due to the small radius of curvature of the
particle's corners).

\renewcommand{\thepage}{S-\arabic{page}}
\setcounter{page}{1} 

\subsection{Simulation details}

Replica-exchange Monte Carlo (REMC) simulations are generally employed to 
enhance the sampling from uneven free-energy 
landscapes~\cite{Lyubartsev92,hukushima96}. The 
technique is based on the definition of an extended ensemble, $Q_{{\rm 
ext}}=\prod_i^{N_r} Q_i$, $Q_i$ being the partition function of ensemble 
$i$. For athermal systems, those composed of hard particles, we employ a 
pressure expansion of the isobaric 
ensemble~\cite{Okabe01,Basurto20}. Thus, $Q_{{\rm 
ext}}=\prod_i^{N_r} Q(N,P_{i},T)$, where $N$, $T$, and $P_{i}$ are the number 
of particles, the temperature, and the 2D-pressure, respectively. Here, all 
$N_r$ ensembles share the same $N$ and $T$, but each one has a different $P_i$. 
Also, we define $N_r$ simulation cells each one placed in a different ensemble. 
Each simulation cell samples a given $NP_iT$ ensemble following a standard MC 
procedure. This is carried out by implementing trials of particle 
displacements, particle rotations, area-changes of the 
simulation cell, and shape changes of the simulation cell. However, the 
definition of $Q_{{\rm ext}}$ allows the inclusion of swap trials. These are 
carried out between simulation boxes placed at ensembles with $P_i$ and 
$P_{i+1}$, with acceptance probability 
$\min\{1,\exp{[\beta(P_{i}- P_{i+1})(A_i-A_{i+1})]}\}$. In this 
expression, $\beta= 1/(k_BT)$, $k_B$ is the Boltzmann constant, and $A_i$ 
and $A_j$ are the areas of replicas $i$ and $j$, 
respectively. We set a geometric progression with 
the replica index for $\beta P_{i}$, from $\beta P_{\rm{min}}$ to 
$\beta P_{\rm{max}}$. The implementation of the simulations mainly works in the 
CPU, each of its cores handling several replicas, and calling the Graphics 
Processing Units (GPUs) for building the neighbors lists (a CUDA-MPI 
implementation). 

\begin{figure}[t!]
\centering
\includegraphics[width=0.95\linewidth]{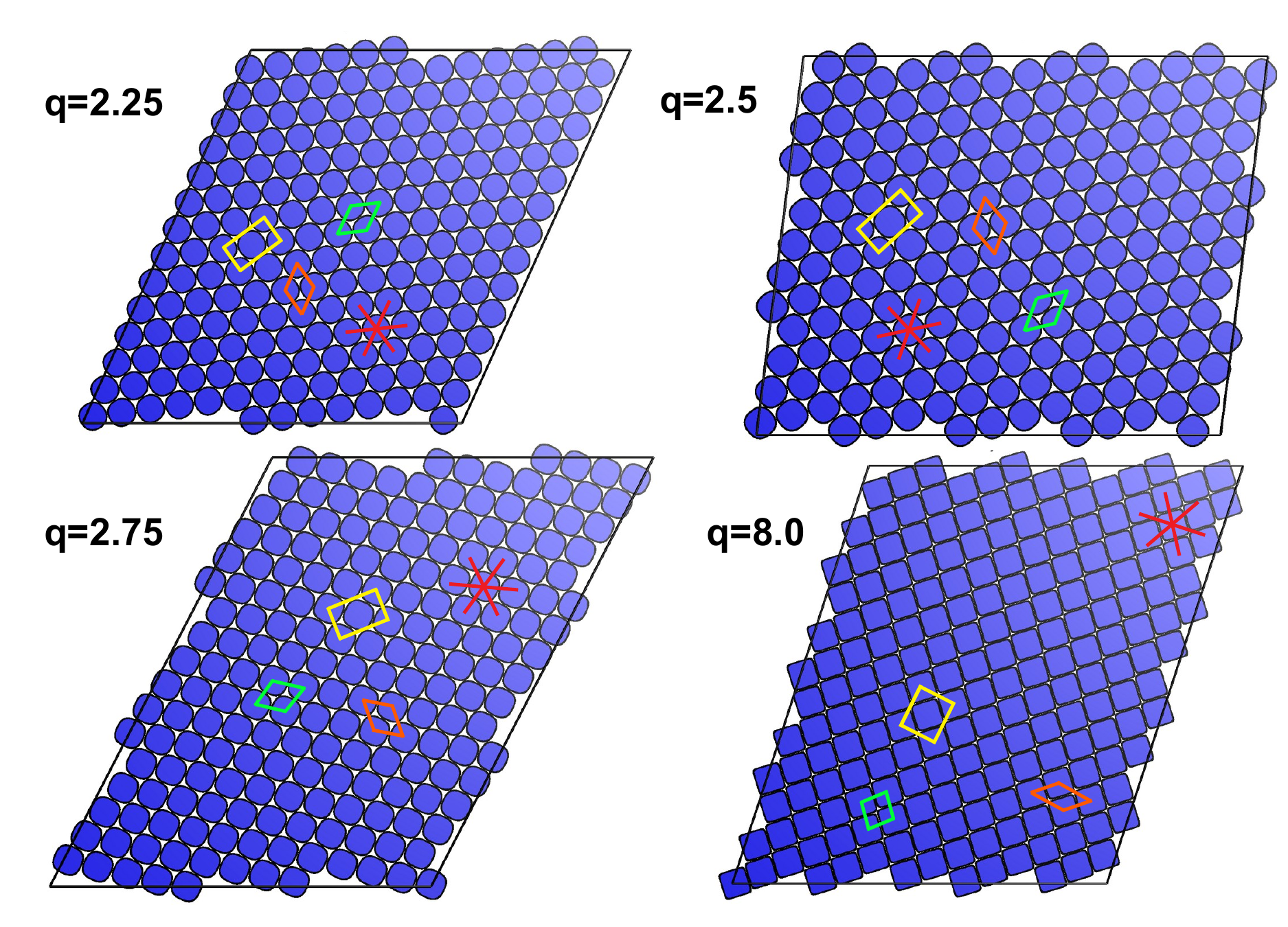}
\caption{Snapshots obtained for $\beta P \sigma^2=10000$ showing the nearly 
close packing structures for different deformation parameters. We have drawn 
a sixfold star to highlight deviations from the hexagonal arrangement. Also, 
we have depicted the conventional unit cell of the centered rectangular 
lattice (in yellow), the rhombic primitive unit cell (in 
green), and the primitive unit cell defined by Jiao et al.~\cite{Jiao08} (in 
red).}
\label{fig:snapshots}
\end{figure}

For the continuous lines shown in panels a) to d) of Figs.~1 and 2 of the 
manuscript, and for Fig.~\ref{fig:case2.5} of this SM, 
we have set $N_r=100$, $N=196$, $\beta P_{\rm{min}} \sigma^2=5$, and $\beta 
P_{\rm{max}} \sigma^2=50$ for all studied $\defp$ values except for $\defp\leq 
3.0$, 
where we are setting $N_r=120$ and $\beta P_{\rm{max}} \sigma^2=200$. To obtain 
high-pressure configurations, we have compressed 
the obtained structures by setting $\beta P_{\rm{min}} \sigma^2=10$ and $\beta 
P_{\rm{min}} \sigma^2=10000$. To confirm the $N=196$ outcomes for $\defp=6.0$, 
we 
have also carried out REMC simulations with $N_r=40$ and $N=6400$ in the 
density region $0.722<\eta<0.825$. Finally, we have also performed 
some $N=6400$ standard Monte Carlo simulations at interesting densities for all 
studied $\defp$ cases to access the quasi-long-range 
(QLR) behavior of the correlation functions defined in the manuscript.  

We start all the simulations with $N=196$ from loose random configurations. We 
have found that starting from a tight square lattice reaches the same 
equilibrium state a little bit faster. Thus, all simulations with $N=6400$ 
particles are started from a square lattice. Once a steady-state is  
achieved, we perform the several averages defined in the manuscript. 

\subsection{Optimal packing structures}

\begin{figure}[t!]
\centering
\includegraphics[width=0.95\linewidth]{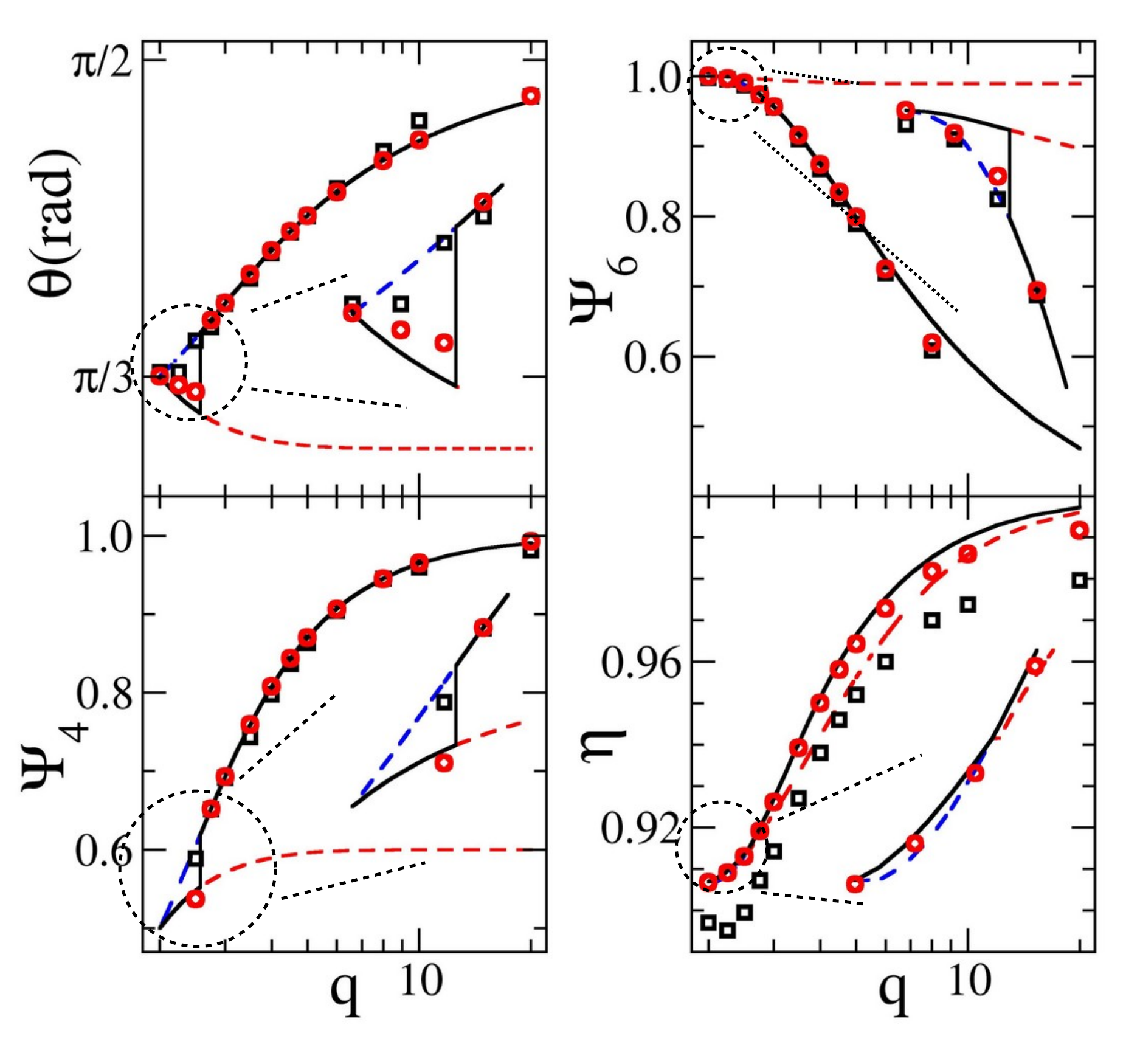}
\caption{Lattice angle of the primitive rhombic unit cell, $\theta$,
  global bond-orientational order parameters, $\Psi_4$, $\Psi_6$, and
  packing fraction $\eta$, as a function of $\defp$. Red and blue
  dashed lines correspond to $\Lambda_0$ and
  $\Lambda_1$, respectively. The continuous black line
  corresponds to the optimal structure. The black symbols are the
  simulation outcomes for $\beta P \sigma^2=200$. The red symbols
  correspond to $\beta P \sigma^2=10000$. Insets of each panel
  zoom in the $2\leq \defp \leq2.8$ region.}
\label{fig:angles}
\end{figure}

There are two conjectured optimal close-packed structures, denoted by 
$\Lambda_0$ and $\Lambda_1$~\cite{Jiao08}, which can maximize the packing 
density for $\defp\geq2$. $\Lambda_0$
maximizes the packing density for $\defp\lesssim2.572$ and $\Lambda_1$
does it for $\defp\gtrsim2.572$. These structures were also
experimentally observed with depletion stabilized silica
superballs~\cite{Rossi15} and hollow silica cubes~\cite{Meijer2019}.
Both of them are centered rectangular Bravais lattices that can be
described by several different unit cells. In Fig~\ref{fig:snapshots} we
highlighted in the snapshots the conventional unit cells, the primitive
rhombic unit cells, and the primitive unit cells defined in
Ref~\cite{Jiao08}, which are parallelograms. The primitive lattice
vectors of the parallelograms are $\vec{e}_1=2\vec{e}_x$ and
$\vec{e}_2=\vec{e}_x+(2^\defp-1)^{1/\defp}\vec{e}_y$ in case of
$\Lambda_0$, whereas
$\vec{e}_1=2^{(1-1/\defp)}\vec{e}_x+2^{(1-1/\defp)}\vec{e}_y$ and
$\vec{e}_2=(2^{-1/\defp}-2^{1/2}s)\vec{e}_x+(2^{-1/\defp}+2^{1/2}s)\vec{e}_y$
in case of $\Lambda_1$. Here, $s$ is the smallest positive root of
$|2^{-(1+1/\defp)}-2^{-1/2}s|^\defp+|2^{-(1+1/\defp)}+2^{-1/2}s|^\defp=1$.
The primitive lattice vectors associated with the rhombic
unit cells are $\vec{a}=\vec{e}_2$ and $\vec{b}=\vec{e}_2-\vec{e}_1$
for both structures. The tilt angle of these rhombic unit cells is given
by
\begin{equation}
  \theta=\arccos(\vec{a} \cdot \vec{b}/|\vec{a}||\vec{b}|)
  \, .
  \label{theta}
\end{equation}
In addition, the maximal packing fraction is
\begin{equation}
\eta_{\mathrm{cp}}=\frac{4}{|\vec{a}||\vec{b}|\sin{\theta}}
\int_0^1(1-x^\defp)^{1/\defp}dx, 
\end{equation}
which gives the boundary of the unreachable region of the phase
diagram indicated by the black line of Fig.~4 of the main text. This line
corresponds to the infinite pressure limit of the isobaric ensemble,
where only the optimal arrangements survive. Therefore, we expect to have 
rhombic solid phases with structures very close to $\Lambda_0$ and 
$\Lambda_1$ at high but finite pressures, which gives us the possibility of 
checking our simulation results in the high-pressure limit.

For this reason, in Fig~\ref{fig:angles} we compare the simulation
data for $\beta P \sigma^2 = 200$ (indicated by black squares) and for
$\beta P \sigma^2 = 10000$ (indicated by red circles) with the
theoretical values of the perfect rhombatic lattices. Red and blue
dashed lines correspond to $\Lambda_0$ and $\Lambda_1$ structures,
respectively. We compare the lattice angle, $\theta$, the global
bond-orientational order parameters, $\Psi_4$ and $\Psi_6$ (defined in
the main text), and the packing fraction, $\eta$. In general, it is
observed how the simulation results approach all theoretical
quantities and that the agreement improves with increasing
pressure. This is particularly true for the packing
fraction. Nonetheless, we observe a tendency for the simulation data
to depart the theoretical maximal packing fraction as the deformation
parameter, $\defp$, increases, pointing out that an even higher
pressure is needed to reach the close-packing limit in case of
square-like particles. More importantly, all simulation results are close and
below the conjectured maximal packing limit, which supports
$\Lambda_0$ and $\Lambda_1$ as optimal arrangements.

A perfect rhombic lattice has
$\Psi_6=|1+e^{\bm{i}6\theta}-e^{\bm{i}3\theta}|/3$ and
$\Psi_4=|1+e^{\bm{i}4\theta}|/2$. These expressions, substituting
$\theta$ from Eq.(\ref{theta}), give the dashed curves of the $\Psi_4$
and $\Psi_6$ panels of Fig.~\ref{fig:angles}. The agreement with the
simulations is good. However, in the limiting cases, there appears
some discrepancies explained below. For the theoretical curves we have
$\Psi_6 \rightarrow 1$ and $\Psi_4 \rightarrow 1/2$ in the disk
limit ($\defp \rightarrow 2$, $\theta \rightarrow \pi/3$),
and $\Psi_6 \rightarrow 1/3$ and $\Psi_4 \rightarrow 1$ in
the square limit ($\defp\rightarrow \infty$, $\theta \rightarrow \pi/2$). These 
limits are not necessarily valid at finite pressure when QLR
positional order replaces the long-range order of the perfect lattice. 
Especially, $\Psi_4$ can be quite different given 
that the four nearest neighbors of a particle turn ill-defined in the disk 
limit. Due to the small fluctuations of the particles' positions and 
orientations, we get $\Psi_4 \approx 0$. A similar problem appears with 
$\Psi_6$ in the square limit.

\begin{figure}[t!]
\centering
\includegraphics[width=0.95\linewidth]{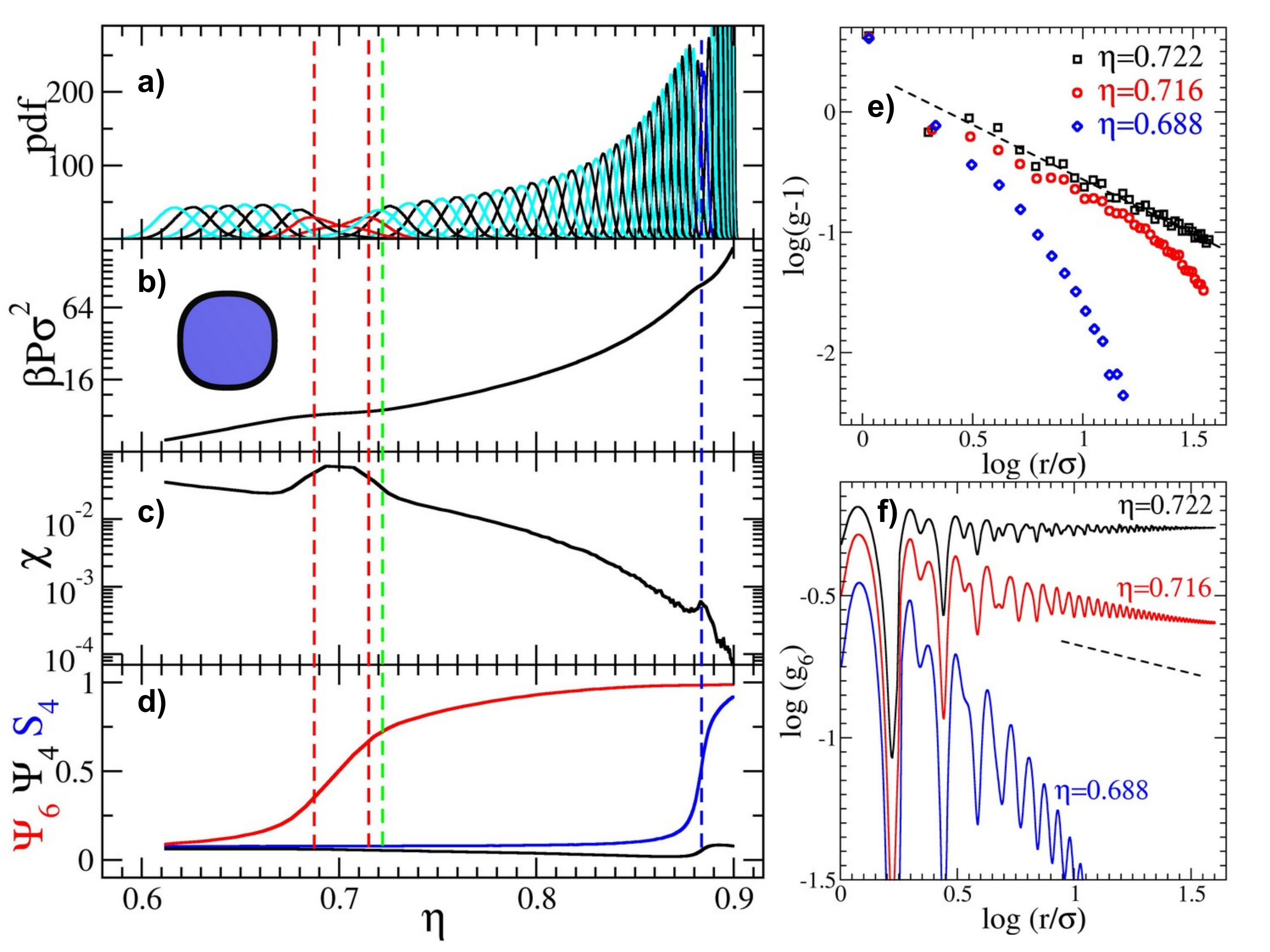}
\caption{a) Probability density functions (red and blue lines are
  employed to highlight the histograms close to the isotropic--hexatic
  and the plastic solid--rhombic solid transitions, respectively), b)
  dimensionless pressure, $\beta P \sigma^2$, c) dimensionless
  isothermal compressibility, $\chi$, and d) global order parameters,
  $\Psi_6$ (red line), $\Psi_4$ (black line), and $S_4$ (blue line),
  as a function of the packing fraction, $\eta$, for a system of
  $N=196$ superdisks with $\defp=2.5$ (the superdisk's shape is shown
  as an inset). The biphasic region of the first-order
  isotropic--hexatic transition is delimited by the vertical dashed
  red lines. The vertical dashed blue and green lines signal the
  development of orientational order and QLR positional order. Panel
  e) shows the peaks of $\log(g-1)$ and panel f) shows $\log(g_6)$ as
  a function of $\log(r/\sigma)$ for a system of $N=6400$
  superdisks. The black dashed line in panel f) depicts a slope of
  $-1/4$.}
\label{fig:case2.5}
\end{figure}

For $\defp<2.75$ the $\beta P \sigma^2 = 10000$ data agree with the
$\Lambda_0$ curves, whereas they are consistent with the $\Lambda_1$
curves for $\defp\geq 2.75$ (see the panels corresponding to $\theta$
and $\Psi_4$). Similarly, the data obtained for not so high pressure
agree with the $\Lambda_0$ curves for $\defp<2.5$ and with the
$\Lambda_1$ curves for $\defp\geq 2.5$. Thus, there appears a
difference between high and very high pressures for the $\defp=2.75$
case (see the inserts in all panels of Fig.~\ref{fig:angles}) that
suggests a transition from $\Lambda_0$-like structures to
$\Lambda_1$-ones. From the probability density functions, we estimate
this transition to be around $\eta = 0.908$. This is why we are
drawing a tilt dashed line splitting the rhombic solid from the
hexagonal solid in the phase diagram (Fig.~3) shown in the manuscript.

\subsection{Results for the $\defp=2.5$ case}

In this section of the SM, we show the outcomes from REMC simulations
for the system with $\defp=2.5$. This case is close to
the disk limit and the melting of these particles corresponds to a
different scenario, differing from the cases presented in the main text. Namely,
there appears a discontinuous two-step melting with a first-order 
transition between the isotropic and the hexatic phases and a continuous 
transition between the hexatic and the plastic solid phase. The first-order
transition can be clearly read from panel a), which shows how the
probability density functions distort from the Gaussian-shape and turn
bimodal from $\eta\approx0.688$ to $\eta\approx0.716$, signaling the
coexistence region between the isotropic and the hexatic phases. Note
that our low-boundary estimation of this region is smaller than the
generally accepted value. This is due to size effects. Furthermore, in
this region, we have a plateau in the EOS, a large peak in $\chi$, and
a sudden increase in $\Psi_6$. We highlight this region with the
vertical red dashed lines crossing from panel a) to d). Also, panel e)
depicts the change from short-range to QLR behavior of the $g(r)$ and
panel f) the change from short-range to long-range of $g_6(r)$, passing through
QLR correlations. We estimate the establishment of long-range and QLR for
$g_6(r)$ and $g(r)$, respectively, occurring at $\eta\approx0.72$. We
draw a vertical cyan dashed line at this density. All these features
are in common with the $\defp=2.0$ case, thus we recover the third
scenario mentioned in the first paragraph of the main text.

\begin{figure}[t!]
\centering
\includegraphics[width=0.95\linewidth]{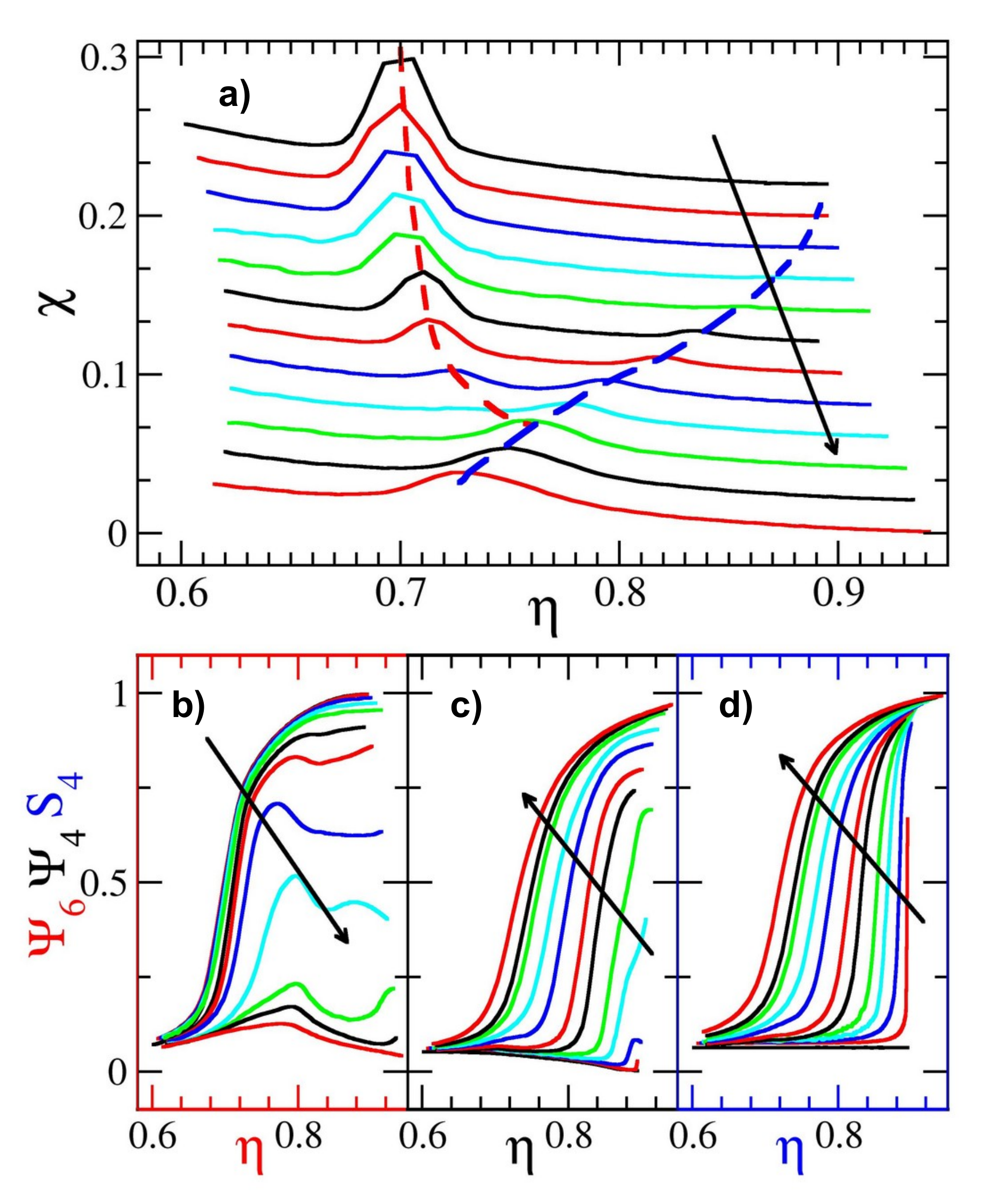}
\caption{a) Dimensionless isothermal compressibility, $\chi$ , b) global 
six-fold bond-orientational order parameter, $\Psi_6$, c) global four-fold
bond-orientational order parameter, $\Psi_4$, and d) global tetratic
particle orientational order parameter, $S_4$, as a function of the
packing fraction, $\eta$, for $\defp=$ 2, 2.25, 2.5, 2.75, 3.0, 3.5,
4.0, 5.0, 6.0, 8.0, 10.0, and 20.0. Arrows indicate the increasing
direction of $\defp$. In panel a) curves are given an offset of
0.02, and red and blue dotted lines are guides to the eye passing
through the compressibility peaks.}
\label{fig:shift}
\end{figure}

Differences between disks and superdisks with $\defp=2.5$ appear
above $\eta \approx0.72$. To begin with, given that particles with
$\defp>2$ break the circular symmetry, the hexagonal solid-phase turns
into a plastic phase with hexagonal crystal structure, also called
hexagonal solid rotator phase. This phase has QLR positional order,
but particle-orientation correlations are short-ranged. The particles
orientational degrees of freedom frozen only at very high pressures and
high densities to yield an orientationally ordered $\Lambda_0$ rhombic
solid phase (see the snapshot in Fig.~\ref{fig:angles}). We detect a
smooth solid--solid transition, where the Gaussian-shape of all PDFs
is preserved, but their height (width) decreases (increases) a little
at it. This is accompanied by a small peak of $\chi$ and a sudden
increase in $S_4$. A vertical dashed line is placed across panels
a)-d) of Fig.~\ref{fig:case2.5} to signal this transition. All these
features resemble the behavior of rounded-hard
squares~\cite{Zhao2012,Avendano12}, hard ellipses~\cite{Bautista14},
and discorectangles~\cite{Bates2000} with quasi-circular symmetry.

\subsection{Dimensionless isothermal compressibility and global order 
parameters}

We present here the curves of the dimensionless isothermal compressibility, 
$\chi= d \rho/d (\beta P) = N (\langle 
\rho^2 \rangle \! - \!\langle \rho \rangle^2)/\langle \rho \rangle^2$, and the
measured global order parameters as obtained from the REMC as a function of the 
deformation parameter, $\defp$. The global order parameters 
are the six- and four-fold bond-orientational and the tetratic 
particle orientational ones. The aim of doing this is two-fold. 
First, we highlight that the REMC technique produces very smooth curves for 
the isothermal compressibility, which allows for the easy detection of the 
phase boundaries. This contrasts with the standard point-by-point simulations. 
Second, we summarize all cases making clear how we build the global phase 
diagram shown in Fig.~3 of the main text.

The trends of the $\chi$ peaks with increasing $\defp$ are depicted in
Fig.~\ref{fig:shift} a). Note that the $\chi$ peak appearing at low
densities shifting to the right correlates with the sudden increase of
$\Psi_6$, whereas the one appearing at high densities correlates with
the increase of $S_4$ and $\Psi_4$. Also, it is observed that
the $S_4$-$\Psi_4$ correlation expands all the $\defp$ interval (see
Fig.~\ref{fig:shift} b) and c)), although the $S_4$ steep increase
anticipates that for $\Psi_4$. We add Y-shaped guides to the eye as
red and blue dashed lines in Fig.~\ref{fig:shift} a), suggesting the
vanishing of the isotropic-hexatic transition. The right-hand side
branch (blue) of the Y corresponds to the building of orientational
order, whereas the left-hand side branch (red) signals the
isotropic-hexatic transition.


\begin{thebibliography}{35}%
\makeatletter
\providecommand \@ifxundefined [1]{%
 \@ifx{#1\undefined}
}%
\providecommand \@ifnum [1]{%
 \ifnum #1\expandafter \@firstoftwo
 \else \expandafter \@secondoftwo
 \fi
}%
\providecommand \@ifx [1]{%
 \ifx #1\expandafter \@firstoftwo
 \else \expandafter \@secondoftwo
 \fi
}%
\providecommand \natexlab [1]{#1}%
\providecommand \enquote  [1]{``#1''}%
\providecommand \bibnamefont  [1]{#1}%
\providecommand \bibfnamefont [1]{#1}%
\providecommand \citenamefont [1]{#1}%
\providecommand \href@noop [0]{\@secondoftwo}%
\providecommand \href [0]{\begingroup \@sanitize@url \@href}%
\providecommand \@href[1]{\@@startlink{#1}\@@href}%
\providecommand \@@href[1]{\endgroup#1\@@endlink}%
\providecommand \@sanitize@url [0]{\catcode `\\12\catcode `\$12\catcode
  `\&12\catcode `\#12\catcode `\^12\catcode `\_12\catcode `\%12\relax}%
\providecommand \@@startlink[1]{}%
\providecommand \@@endlink[0]{}%
\providecommand \url  [0]{\begingroup\@sanitize@url \@url }%
\providecommand \@url [1]{\endgroup\@href {#1}{\urlprefix }}%
\providecommand \urlprefix  [0]{URL }%
\providecommand \Eprint [0]{\href }%
\providecommand \doibase [0]{http://dx.doi.org/}%
\providecommand \selectlanguage [0]{\@gobble}%
\providecommand \bibinfo  [0]{\@secondoftwo}%
\providecommand \bibfield  [0]{\@secondoftwo}%
\providecommand \translation [1]{[#1]}%
\providecommand \BibitemOpen [0]{}%
\providecommand \bibitemStop [0]{}%
\providecommand \bibitemNoStop [0]{.\EOS\space}%
\providecommand \EOS [0]{\spacefactor3000\relax}%
\providecommand \BibitemShut  [1]{\csname bibitem#1\endcsname}%
\let\auto@bib@innerbib\@empty
\bibitem [{\citenamefont {Murray}\ and\ \citenamefont
  {Van~Winkle}(1987)}]{Murray87}%
  \BibitemOpen
  \bibfield  {author} {\bibinfo {author} {\bibfnamefont {C.~A.}\ \bibnamefont
  {Murray}}\ and\ \bibinfo {author} {\bibfnamefont {D.~H.}\ \bibnamefont
  {Van~Winkle}},\ }\href {\doibase 10.1103/PhysRevLett.58.1200} {\bibfield
  {journal} {\bibinfo  {journal} {Phys. Rev. Lett.}\ }\textbf {\bibinfo
  {volume} {58}},\ \bibinfo {pages} {1200} (\bibinfo {year}
  {1987})}\BibitemShut {NoStop}%
\bibitem [{\citenamefont {Donev}\ \emph {et~al.}(2006)\citenamefont {Donev},
  \citenamefont {Burton}, \citenamefont {Stillinger},\ and\ \citenamefont
  {Torquato}}]{Donev06}%
  \BibitemOpen
  \bibfield  {author} {\bibinfo {author} {\bibfnamefont {A.}~\bibnamefont
  {Donev}}, \bibinfo {author} {\bibfnamefont {J.}~\bibnamefont {Burton}},
  \bibinfo {author} {\bibfnamefont {F.~H.}\ \bibnamefont {Stillinger}}, \ and\
  \bibinfo {author} {\bibfnamefont {S.}~\bibnamefont {Torquato}},\ }\href@noop
  {} {\bibfield  {journal} {\bibinfo  {journal} {Phys. Rev. B}\ }\textbf
  {\bibinfo {volume} {73}},\ \bibinfo {pages} {054109} (\bibinfo {year}
  {2006})}\BibitemShut {NoStop}%
\bibitem [{\citenamefont {Han}\ \emph {et~al.}(2008)\citenamefont {Han},
  \citenamefont {Ha}, \citenamefont {Alsayed},\ and\ \citenamefont
  {Yodh}}]{Han-Ha-Alsayed-Yodh_PRE_2008}%
  \BibitemOpen
  \bibfield  {author} {\bibinfo {author} {\bibfnamefont {Y.}~\bibnamefont
  {Han}}, \bibinfo {author} {\bibfnamefont {N.~Y.}\ \bibnamefont {Ha}},
  \bibinfo {author} {\bibfnamefont {A.~M.}\ \bibnamefont {Alsayed}}, \ and\
  \bibinfo {author} {\bibfnamefont {A.~G.}\ \bibnamefont {Yodh}},\ }\href
  {\doibase 10.1103/PhysRevE.77.041406} {\bibfield  {journal} {\bibinfo
  {journal} {Phys. Rev. E}\ }\textbf {\bibinfo {volume} {77}},\ \bibinfo
  {pages} {041406} (\bibinfo {year} {2008})}\BibitemShut {NoStop}%
\bibitem [{\citenamefont {Halperin}\ and\ \citenamefont
  {Nelson}(1978)}]{Halperin-Nelson}%
  \BibitemOpen
  \bibfield  {author} {\bibinfo {author} {\bibfnamefont {B.~I.}\ \bibnamefont
  {Halperin}}\ and\ \bibinfo {author} {\bibfnamefont {D.~R.}\ \bibnamefont
  {Nelson}},\ }\href@noop {} {\bibfield  {journal} {\bibinfo  {journal} {Phys.
  Rev. Lett.}\ }\textbf {\bibinfo {volume} {41}},\ \bibinfo {pages} {121}
  (\bibinfo {year} {1978})}\BibitemShut {NoStop}%
\bibitem [{\citenamefont {Kosterlitz}\ and\ \citenamefont
  {Thouless}(1973)}]{Kosterlitz-Thouless}%
  \BibitemOpen
  \bibfield  {author} {\bibinfo {author} {\bibfnamefont {J.~M.}\ \bibnamefont
  {Kosterlitz}}\ and\ \bibinfo {author} {\bibfnamefont {D.~J.}\ \bibnamefont
  {Thouless}},\ }\href@noop {} {\bibfield  {journal} {\bibinfo  {journal} {J.
  Phys. C}\ }\textbf {\bibinfo {volume} {6}},\ \bibinfo {pages} {1181}
  (\bibinfo {year} {1973})}\BibitemShut {NoStop}%
\bibitem [{\citenamefont {Young}(1979)}]{Young_PRB_1979}%
  \BibitemOpen
  \bibfield  {author} {\bibinfo {author} {\bibfnamefont {A.~P.}\ \bibnamefont
  {Young}},\ }\href {\doibase 10.1103/PhysRevB.19.1855} {\bibfield  {journal}
  {\bibinfo  {journal} {Phys. Rev. B}\ }\textbf {\bibinfo {volume} {19}},\
  \bibinfo {pages} {1855} (\bibinfo {year} {1979})}\BibitemShut {NoStop}%
\bibitem [{\citenamefont {Anderson}\ \emph {et~al.}(2017)\citenamefont
  {Anderson}, \citenamefont {Antonaglia}, \citenamefont {Millan}, \citenamefont
  {Engel},\ and\ \citenamefont {Glotzer}}]{Anderson2017}%
  \BibitemOpen
  \bibfield  {author} {\bibinfo {author} {\bibfnamefont {J.}~\bibnamefont
  {Anderson}}, \bibinfo {author} {\bibfnamefont {J.}~\bibnamefont
  {Antonaglia}}, \bibinfo {author} {\bibfnamefont {J.}~\bibnamefont {Millan}},
  \bibinfo {author} {\bibfnamefont {M.}~\bibnamefont {Engel}}, \ and\ \bibinfo
  {author} {\bibfnamefont {S.}~\bibnamefont {Glotzer}},\ }\href@noop {}
  {\bibfield  {journal} {\bibinfo  {journal} {Phys. Rev. X}\ }\textbf {\bibinfo
  {volume} {7}},\ \bibinfo {pages} {021001} (\bibinfo {year}
  {2017})}\BibitemShut {NoStop}%
\bibitem [{\citenamefont {Karnchanaphanurach}\ \emph
  {et~al.}(2000)\citenamefont {Karnchanaphanurach}, \citenamefont {Lin},\ and\
  \citenamefont {Rice}}]{Karnchanaphanurach-Lin-Rice_PRE_2000}%
  \BibitemOpen
  \bibfield  {author} {\bibinfo {author} {\bibfnamefont {P.}~\bibnamefont
  {Karnchanaphanurach}}, \bibinfo {author} {\bibfnamefont {B.}~\bibnamefont
  {Lin}}, \ and\ \bibinfo {author} {\bibfnamefont {S.~A.}\ \bibnamefont
  {Rice}},\ }\href {\doibase 10.1103/PhysRevE.61.4036} {\bibfield  {journal}
  {\bibinfo  {journal} {Phys. Rev. E}\ }\textbf {\bibinfo {volume} {61}},\
  \bibinfo {pages} {4036} (\bibinfo {year} {2000})}\BibitemShut {NoStop}%
\bibitem [{\citenamefont {Marcus}\ and\ \citenamefont
  {Rice}(1996)}]{Marcus-Rice_PRL_1996}%
  \BibitemOpen
  \bibfield  {author} {\bibinfo {author} {\bibfnamefont {A.~H.}\ \bibnamefont
  {Marcus}}\ and\ \bibinfo {author} {\bibfnamefont {S.~A.}\ \bibnamefont
  {Rice}},\ }\href {\doibase 10.1103/PhysRevLett.77.2577} {\bibfield  {journal}
  {\bibinfo  {journal} {Phys. Rev. Lett.}\ }\textbf {\bibinfo {volume} {77}},\
  \bibinfo {pages} {2577} (\bibinfo {year} {1996})}\BibitemShut {NoStop}%
\bibitem [{\citenamefont {Qi}\ \emph {et~al.}(2014)\citenamefont {Qi},
  \citenamefont {Gantapara},\ and\ \citenamefont
  {Dijkstra}}]{Dijkstra_SOFTM_2014}%
  \BibitemOpen
  \bibfield  {author} {\bibinfo {author} {\bibfnamefont {W.}~\bibnamefont
  {Qi}}, \bibinfo {author} {\bibfnamefont {A.~P.}\ \bibnamefont {Gantapara}}, \
  and\ \bibinfo {author} {\bibfnamefont {M.}~\bibnamefont {Dijkstra}},\ }\href
  {\doibase {10.1039/c4sm00125g}} {\bibfield  {journal} {\bibinfo  {journal}
  {{Soft Matter}}\ }\textbf {\bibinfo {volume} {{10}}},\ \bibinfo {pages}
  {5449} (\bibinfo {year} {{2014}})}\BibitemShut {NoStop}%
\bibitem [{\citenamefont {Thorneywork}\ \emph {et~al.}(2017)\citenamefont
  {Thorneywork}, \citenamefont {Abbott}, \citenamefont {Aarts},\ and\
  \citenamefont {Dullens}}]{Thorneywork17}%
  \BibitemOpen
  \bibfield  {author} {\bibinfo {author} {\bibfnamefont {A.}~\bibnamefont
  {Thorneywork}}, \bibinfo {author} {\bibfnamefont {J.}~\bibnamefont {Abbott}},
  \bibinfo {author} {\bibfnamefont {D.}~\bibnamefont {Aarts}}, \ and\ \bibinfo
  {author} {\bibfnamefont {R.}~\bibnamefont {Dullens}},\ }\href@noop {}
  {\bibfield  {journal} {\bibinfo  {journal} {Phys. Rev. Lett.}\ }\textbf
  {\bibinfo {volume} {118}},\ \bibinfo {pages} {158001} (\bibinfo {year}
  {2017})}\BibitemShut {NoStop}%
\bibitem [{\citenamefont {Wojciechowski}\ and\ \citenamefont
  {Frenkel}(2004)}]{Wojciechowski04}%
  \BibitemOpen
  \bibfield  {author} {\bibinfo {author} {\bibfnamefont {K.~W.}\ \bibnamefont
  {Wojciechowski}}\ and\ \bibinfo {author} {\bibfnamefont {D.}~\bibnamefont
  {Frenkel}},\ }\href@noop {} {\bibfield  {journal} {\bibinfo  {journal} {Comp.
  Met. Sci. Technol.}\ }\textbf {\bibinfo {volume} {10}},\ \bibinfo {pages}
  {235} (\bibinfo {year} {2004})}\BibitemShut {NoStop}%
\bibitem [{\citenamefont {Schilling}\ \emph {et~al.}(2005)\citenamefont
  {Schilling}, \citenamefont {Pronk}, \citenamefont {Mulder},\ and\
  \citenamefont {Frenkel}}]{Schilling05}%
  \BibitemOpen
  \bibfield  {author} {\bibinfo {author} {\bibfnamefont {T.}~\bibnamefont
  {Schilling}}, \bibinfo {author} {\bibfnamefont {S.}~\bibnamefont {Pronk}},
  \bibinfo {author} {\bibfnamefont {B.}~\bibnamefont {Mulder}}, \ and\ \bibinfo
  {author} {\bibfnamefont {D.}~\bibnamefont {Frenkel}},\ }\href@noop {}
  {\bibfield  {journal} {\bibinfo  {journal} {Phys. Rev. E.}\ }\textbf
  {\bibinfo {volume} {71}},\ \bibinfo {pages} {036138} (\bibinfo {year}
  {2005})}\BibitemShut {NoStop}%
\bibitem [{\citenamefont {Bernard}\ and\ \citenamefont
  {Krauth}(2011)}]{Bernard_PRL_2011}%
  \BibitemOpen
  \bibfield  {author} {\bibinfo {author} {\bibfnamefont {E.~P.}\ \bibnamefont
  {Bernard}}\ and\ \bibinfo {author} {\bibfnamefont {W.}~\bibnamefont
  {Krauth}},\ }\href {\doibase {10.1103/PhysRevLett.107.155704}} {\bibfield
  {journal} {\bibinfo  {journal} {{Phys. Rev. Lett.}}\ }\textbf {\bibinfo
  {volume} {{107}}},\ \bibinfo {pages} {155704} (\bibinfo {year}
  {{2011}})}\BibitemShut {NoStop}%
\bibitem [{\citenamefont {Engel}\ \emph {et~al.}(2013)\citenamefont {Engel},
  \citenamefont {Anderson}, \citenamefont {Glotzer}, \citenamefont {Isobe},
  \citenamefont {Bernard},\ and\ \citenamefont {Krauth}}]{Engel_PRE_2013}%
  \BibitemOpen
  \bibfield  {author} {\bibinfo {author} {\bibfnamefont {M.}~\bibnamefont
  {Engel}}, \bibinfo {author} {\bibfnamefont {J.~A.}\ \bibnamefont {Anderson}},
  \bibinfo {author} {\bibfnamefont {S.~C.}\ \bibnamefont {Glotzer}}, \bibinfo
  {author} {\bibfnamefont {M.}~\bibnamefont {Isobe}}, \bibinfo {author}
  {\bibfnamefont {E.~P.}\ \bibnamefont {Bernard}}, \ and\ \bibinfo {author}
  {\bibfnamefont {W.}~\bibnamefont {Krauth}},\ }\href {\doibase
  {10.1103/PhysRevE.87.042134}} {\bibfield  {journal} {\bibinfo  {journal}
  {{Phys. Rev. E}}\ }\textbf {\bibinfo {volume} {{87}}},\ \bibinfo {pages}
  {042134} (\bibinfo {year} {{2013}})}\BibitemShut {NoStop}%
\bibitem [{\citenamefont {Kapfer}\ and\ \citenamefont
  {Krauth}(2015)}]{Kapfer2015}%
  \BibitemOpen
  \bibfield  {author} {\bibinfo {author} {\bibfnamefont {S.}~\bibnamefont
  {Kapfer}}\ and\ \bibinfo {author} {\bibfnamefont {W.}~\bibnamefont
  {Krauth}},\ }\href@noop {} {\bibfield  {journal} {\bibinfo  {journal} {Phys.
  Rev. Lett}\ }\textbf {\bibinfo {volume} {114}},\ \bibinfo {pages} {035702}
  (\bibinfo {year} {2015})}\BibitemShut {NoStop}%
\bibitem [{\citenamefont {Li}\ and\ \citenamefont
  {Ciamarra}(2020)}]{Li-Ciamarra_PRL_2020}%
  \BibitemOpen
  \bibfield  {author} {\bibinfo {author} {\bibfnamefont {Y.-W.}\ \bibnamefont
  {Li}}\ and\ \bibinfo {author} {\bibfnamefont {M.~P.}\ \bibnamefont
  {Ciamarra}},\ }\href {\doibase 10.1103/PhysRevLett.124.218002} {\bibfield
  {journal} {\bibinfo  {journal} {Phys. Rev. Lett.}\ }\textbf {\bibinfo
  {volume} {124}},\ \bibinfo {pages} {218002} (\bibinfo {year}
  {2020})}\BibitemShut {NoStop}%
\bibitem [{\citenamefont {Lyubartsev}\ \emph {et~al.}(1992)\citenamefont
  {Lyubartsev}, \citenamefont {Martinovski}, \citenamefont {Shevkunov},\ and\
  \citenamefont {Vorontsov-Velyaminov}}]{Lyubartsev92}%
  \BibitemOpen
  \bibfield  {author} {\bibinfo {author} {\bibfnamefont {A.~P.}\ \bibnamefont
  {Lyubartsev}}, \bibinfo {author} {\bibfnamefont {A.~A.}\ \bibnamefont
  {Martinovski}}, \bibinfo {author} {\bibfnamefont {S.~V.}\ \bibnamefont
  {Shevkunov}}, \ and\ \bibinfo {author} {\bibfnamefont {P.~N.}\ \bibnamefont
  {Vorontsov-Velyaminov}},\ }\href@noop {} {\bibfield  {journal} {\bibinfo
  {journal} {J. Chem. Phys.}\ }\textbf {\bibinfo {volume} {96}},\ \bibinfo
  {pages} {1776} (\bibinfo {year} {1992})}\BibitemShut {NoStop}%
\bibitem [{\citenamefont {Hukushima}\ and\ \citenamefont
  {Nemoto}(1996)}]{hukushima96}%
  \BibitemOpen
  \bibfield  {author} {\bibinfo {author} {\bibfnamefont {K.}~\bibnamefont
  {Hukushima}}\ and\ \bibinfo {author} {\bibfnamefont {K.}~\bibnamefont
  {Nemoto}},\ }\href@noop {} {\bibfield  {journal} {\bibinfo  {journal} {J.
  Phys. Soc. Jpn.}\ }\textbf {\bibinfo {volume} {65}},\ \bibinfo {pages} {1604}
  (\bibinfo {year} {1996})}\BibitemShut {NoStop}%
\bibitem [{\citenamefont {Okabe}\ \emph {et~al.}(2001)\citenamefont {Okabe},
  \citenamefont {Kawata}, \citenamefont {Okamoto},\ and\ \citenamefont
  {Mikami}}]{Okabe01}%
  \BibitemOpen
  \bibfield  {author} {\bibinfo {author} {\bibfnamefont {T.}~\bibnamefont
  {Okabe}}, \bibinfo {author} {\bibfnamefont {M.}~\bibnamefont {Kawata}},
  \bibinfo {author} {\bibfnamefont {Y.}~\bibnamefont {Okamoto}}, \ and\
  \bibinfo {author} {\bibfnamefont {M.}~\bibnamefont {Mikami}},\ }\href@noop {}
  {\bibfield  {journal} {\bibinfo  {journal} {Chem. Phys. Lett.}\ }\textbf
  {\bibinfo {volume} {335}},\ \bibinfo {pages} {435} (\bibinfo {year}
  {2001})}\BibitemShut {NoStop}%
\bibitem [{\citenamefont {Basurto}\ \emph {et~al.}(2018)\citenamefont
  {Basurto}, \citenamefont {Haro-Pérez}, \citenamefont {Vargas},\ and\
  \citenamefont {Odriozola}}]{Gerardo_PCCP_2018}%
  \BibitemOpen
  \bibfield  {author} {\bibinfo {author} {\bibfnamefont {E.}~\bibnamefont
  {Basurto}}, \bibinfo {author} {\bibfnamefont {C.}~\bibnamefont
  {Haro-Pérez}}, \bibinfo {author} {\bibfnamefont {C.~A.}\ \bibnamefont
  {Vargas}}, \ and\ \bibinfo {author} {\bibfnamefont {G.}~\bibnamefont
  {Odriozola}},\ }\href {\doibase 10.1039/C8CP03727B} {\bibfield  {journal}
  {\bibinfo  {journal} {Phys. Chem. Chem. Phys.}\ }\textbf {\bibinfo {volume}
  {20}},\ \bibinfo {pages} {27490} (\bibinfo {year} {2018})}\BibitemShut
  {NoStop}%
\bibitem [{\citenamefont {Avendano}\ and\ \citenamefont
  {Escobedo}(2012)}]{Avendano12}%
  \BibitemOpen
  \bibfield  {author} {\bibinfo {author} {\bibfnamefont {C.}~\bibnamefont
  {Avendano}}\ and\ \bibinfo {author} {\bibfnamefont {F.~A.}\ \bibnamefont
  {Escobedo}},\ }\href@noop {} {\bibfield  {journal} {\bibinfo  {journal} {Soft
  Matter}\ }\textbf {\bibinfo {volume} {8}},\ \bibinfo {pages} {4675} (\bibinfo
  {year} {2012})}\BibitemShut {NoStop}%
\bibitem [{\citenamefont {Zhaglin}\ \emph {et~al.}(2018)\citenamefont
  {Zhaglin}, \citenamefont {Ju}, \citenamefont {Zong}, \citenamefont {Ye},\
  and\ \citenamefont {Zhao}}]{Zhaglin_ChinPhysB_2018}%
  \BibitemOpen
  \bibfield  {author} {\bibinfo {author} {\bibfnamefont {H.}~\bibnamefont
  {Zhaglin}}, \bibinfo {author} {\bibfnamefont {Y.}~\bibnamefont {Ju}},
  \bibinfo {author} {\bibfnamefont {Y.-w.}\ \bibnamefont {Zong}}, \bibinfo
  {author} {\bibfnamefont {F.-f.}\ \bibnamefont {Ye}}, \ and\ \bibinfo {author}
  {\bibfnamefont {K.}~\bibnamefont {Zhao}},\ }\href {\doibase
  10.1088/1674-1056/27/8/088203} {\bibfield  {journal} {\bibinfo  {journal}
  {Chinese Physics B}\ }\textbf {\bibinfo {volume} {27}},\ \bibinfo {pages}
  {088203} (\bibinfo {year} {2018})}\BibitemShut {NoStop}%
\bibitem [{\citenamefont {Zhao}\ \emph {et~al.}(2011)\citenamefont {Zhao},
  \citenamefont {Bruinsma},\ and\ \citenamefont {Mason}}]{Zhao2011}%
  \BibitemOpen
  \bibfield  {author} {\bibinfo {author} {\bibfnamefont {K.}~\bibnamefont
  {Zhao}}, \bibinfo {author} {\bibfnamefont {R.}~\bibnamefont {Bruinsma}}, \
  and\ \bibinfo {author} {\bibfnamefont {T.}~\bibnamefont {Mason}},\
  }\href@noop {} {\bibfield  {journal} {\bibinfo  {journal} {PNAS}\ }\textbf
  {\bibinfo {volume} {108}},\ \bibinfo {pages} {2684} (\bibinfo {year}
  {2011})}\BibitemShut {NoStop}%
\bibitem [{\citenamefont {Jiao}\ \emph {et~al.}(2008)\citenamefont {Jiao},
  \citenamefont {Stillinger},\ and\ \citenamefont {Torquato}}]{Jiao08}%
  \BibitemOpen
  \bibfield  {author} {\bibinfo {author} {\bibfnamefont {Y.}~\bibnamefont
  {Jiao}}, \bibinfo {author} {\bibfnamefont {F.}~\bibnamefont {Stillinger}}, \
  and\ \bibinfo {author} {\bibfnamefont {S.}~\bibnamefont {Torquato}},\
  }\href@noop {} {\bibfield  {journal} {\bibinfo  {journal} {Phys. Rev. Lett.}\
  }\textbf {\bibinfo {volume} {100}},\ \bibinfo {pages} {245504} (\bibinfo
  {year} {2008})}\BibitemShut {NoStop}%
\bibitem [{\citenamefont {Meijer}\ \emph {et~al.}(2017)\citenamefont {Meijer},
  \citenamefont {Pal}, \citenamefont {Ouhajji}, \citenamefont {Lekkerkerker},
  \citenamefont {Philipse},\ and\ \citenamefont
  {Petukhov}}]{Meijer-et.al_NAT.COMM_2017}%
  \BibitemOpen
  \bibfield  {author} {\bibinfo {author} {\bibfnamefont {J.-M.}\ \bibnamefont
  {Meijer}}, \bibinfo {author} {\bibfnamefont {A.}~\bibnamefont {Pal}},
  \bibinfo {author} {\bibfnamefont {S.}~\bibnamefont {Ouhajji}}, \bibinfo
  {author} {\bibfnamefont {H.~N.~W.}\ \bibnamefont {Lekkerkerker}}, \bibinfo
  {author} {\bibfnamefont {A.~P.}\ \bibnamefont {Philipse}}, \ and\ \bibinfo
  {author} {\bibfnamefont {A.~V.}\ \bibnamefont {Petukhov}},\ }\href {\doibase
  10.1038/ncomms14352} {\bibfield  {journal} {\bibinfo  {journal} {Nat.
  Commun.}\ }\textbf {\bibinfo {volume} {8}},\ \bibinfo {pages} {14352}
  (\bibinfo {year} {2017})}\BibitemShut {NoStop}%
\bibitem [{\citenamefont {Rossi}\ \emph
  {et~al.}(2015{\natexlab{a}})\citenamefont {Rossi}, \citenamefont {Soni},
  \citenamefont {Ashton}, \citenamefont {Pine}, \citenamefont {Philipse},
  \citenamefont {Chaikin}, \citenamefont {Dijkstra}, \citenamefont {Sacanna},\
  and\ \citenamefont {Irvine}}]{Rossi5286}%
  \BibitemOpen
  \bibfield  {author} {\bibinfo {author} {\bibfnamefont {L.}~\bibnamefont
  {Rossi}}, \bibinfo {author} {\bibfnamefont {V.}~\bibnamefont {Soni}},
  \bibinfo {author} {\bibfnamefont {D.~J.}\ \bibnamefont {Ashton}}, \bibinfo
  {author} {\bibfnamefont {D.~J.}\ \bibnamefont {Pine}}, \bibinfo {author}
  {\bibfnamefont {A.~P.}\ \bibnamefont {Philipse}}, \bibinfo {author}
  {\bibfnamefont {P.~M.}\ \bibnamefont {Chaikin}}, \bibinfo {author}
  {\bibfnamefont {M.}~\bibnamefont {Dijkstra}}, \bibinfo {author}
  {\bibfnamefont {S.}~\bibnamefont {Sacanna}}, \ and\ \bibinfo {author}
  {\bibfnamefont {W.~T.~M.}\ \bibnamefont {Irvine}},\ }\href {\doibase
  10.1073/pnas.1415467112} {\bibfield  {journal} {\bibinfo  {journal}
  {Proceedings of the National Academy of Sciences}\ }\textbf {\bibinfo
  {volume} {112}},\ \bibinfo {pages} {5286} (\bibinfo {year}
  {2015}{\natexlab{a}})}\BibitemShut {NoStop}%
\bibitem [{\citenamefont {Meijer}\ \emph
  {et~al.}(2019{\natexlab{a}})\citenamefont {Meijer}, \citenamefont {Meester},
  \citenamefont {Hagemans}, \citenamefont {Lekkerkerker}, \citenamefont
  {Philipse},\ and\ \citenamefont {Petukhov}}]{Meijer-et.al_LANGMUIR_2019}%
  \BibitemOpen
  \bibfield  {author} {\bibinfo {author} {\bibfnamefont {J.-M.}\ \bibnamefont
  {Meijer}}, \bibinfo {author} {\bibfnamefont {V.}~\bibnamefont {Meester}},
  \bibinfo {author} {\bibfnamefont {F.}~\bibnamefont {Hagemans}}, \bibinfo
  {author} {\bibfnamefont {H.}~\bibnamefont {Lekkerkerker}}, \bibinfo {author}
  {\bibfnamefont {A.~P.}\ \bibnamefont {Philipse}}, \ and\ \bibinfo {author}
  {\bibfnamefont {A.~V.}\ \bibnamefont {Petukhov}},\ }\href {\doibase
  10.1021/acs.langmuir.8b04330} {\bibfield  {journal} {\bibinfo  {journal}
  {Langmuir}\ }\textbf {\bibinfo {volume} {35}},\ \bibinfo {pages} {4946}
  (\bibinfo {year} {2019}{\natexlab{a}})}\BibitemShut {NoStop}%
\bibitem [{\citenamefont {Mizani}\ \emph {et~al.}(2020)\citenamefont {Mizani},
  \citenamefont {Gurin}, \citenamefont {Aliabadi}, \citenamefont {Salehi},\
  and\ \citenamefont {Varga}}]{Mizani20}%
  \BibitemOpen
  \bibfield  {author} {\bibinfo {author} {\bibfnamefont {S.}~\bibnamefont
  {Mizani}}, \bibinfo {author} {\bibfnamefont {P.}~\bibnamefont {Gurin}},
  \bibinfo {author} {\bibfnamefont {R.}~\bibnamefont {Aliabadi}}, \bibinfo
  {author} {\bibfnamefont {H.}~\bibnamefont {Salehi}}, \ and\ \bibinfo {author}
  {\bibfnamefont {S.}~\bibnamefont {Varga}},\ }\href@noop {} {\bibfield
  {journal} {\bibinfo  {journal} {J. Chem. Phys.}\ }\textbf {\bibinfo {volume}
  {153}},\ \bibinfo {pages} {034501} (\bibinfo {year} {2020})}\BibitemShut
  {NoStop}%
\bibitem [{\citenamefont {Basurto}\ \emph {et~al.}(2020)\citenamefont
  {Basurto}, \citenamefont {Gurin}, \citenamefont {Varga},\ and\ \citenamefont
  {Odriozola}}]{Basurto20}%
  \BibitemOpen
  \bibfield  {author} {\bibinfo {author} {\bibfnamefont {E.}~\bibnamefont
  {Basurto}}, \bibinfo {author} {\bibfnamefont {P.}~\bibnamefont {Gurin}},
  \bibinfo {author} {\bibfnamefont {S.}~\bibnamefont {Varga}}, \ and\ \bibinfo
  {author} {\bibfnamefont {G.}~\bibnamefont {Odriozola}},\ }\href@noop {}
  {\bibfield  {journal} {\bibinfo  {journal} {Phys. Rev. Res.}\ }\textbf
  {\bibinfo {volume} {2}},\ \bibinfo {pages} {013356} (\bibinfo {year}
  {2020})}\BibitemShut {NoStop}%
\bibitem [{\citenamefont {Rossi}\ \emph
  {et~al.}(2015{\natexlab{b}})\citenamefont {Rossi}, \citenamefont {Soni},
  \citenamefont {Ashton}, \citenamefont {Pine}, \citenamefont {Philipse},
  \citenamefont {Chaikin}, \citenamefont {Dijkstra}, \citenamefont {Sacanna},\
  and\ \citenamefont {Irvine}}]{Rossi15}%
  \BibitemOpen
  \bibfield  {author} {\bibinfo {author} {\bibfnamefont {L.}~\bibnamefont
  {Rossi}}, \bibinfo {author} {\bibfnamefont {V.}~\bibnamefont {Soni}},
  \bibinfo {author} {\bibfnamefont {D.}~\bibnamefont {Ashton}}, \bibinfo
  {author} {\bibfnamefont {D.}~\bibnamefont {Pine}}, \bibinfo {author}
  {\bibfnamefont {A.}~\bibnamefont {Philipse}}, \bibinfo {author}
  {\bibfnamefont {P.~M.}\ \bibnamefont {Chaikin}}, \bibinfo {author}
  {\bibfnamefont {M.}~\bibnamefont {Dijkstra}}, \bibinfo {author}
  {\bibfnamefont {S.}~\bibnamefont {Sacanna}}, \ and\ \bibinfo {author}
  {\bibfnamefont {W.}~\bibnamefont {Irvine}},\ }\href@noop {} {\bibfield
  {journal} {\bibinfo  {journal} {PNAS}\ }\textbf {\bibinfo {volume} {112}},\
  \bibinfo {pages} {5286–5290} (\bibinfo {year}
  {2015}{\natexlab{b}})}\BibitemShut {NoStop}%
\bibitem [{\citenamefont {Meijer}\ \emph
  {et~al.}(2019{\natexlab{b}})\citenamefont {Meijer}, \citenamefont {Meester},
  \citenamefont {Hagemans}, \citenamefont {Lekkerkerker}, \citenamefont
  {Philipse},\ and\ \citenamefont {Petukhov}}]{Meijer2019}%
  \BibitemOpen
  \bibfield  {author} {\bibinfo {author} {\bibfnamefont {J.}~\bibnamefont
  {Meijer}}, \bibinfo {author} {\bibfnamefont {V.}~\bibnamefont {Meester}},
  \bibinfo {author} {\bibfnamefont {F.}~\bibnamefont {Hagemans}}, \bibinfo
  {author} {\bibfnamefont {H.}~\bibnamefont {Lekkerkerker}}, \bibinfo {author}
  {\bibfnamefont {A.}~\bibnamefont {Philipse}}, \ and\ \bibinfo {author}
  {\bibfnamefont {A.}~\bibnamefont {Petukhov}},\ }\href@noop {} {\bibfield
  {journal} {\bibinfo  {journal} {Langmuir}\ }\textbf {\bibinfo {volume}
  {35}},\ \bibinfo {pages} {4946} (\bibinfo {year}
  {2019}{\natexlab{b}})}\BibitemShut {NoStop}%
\bibitem [{\citenamefont {Zhao}\ \emph {et~al.}(2012)\citenamefont {Zhao},
  \citenamefont {Bruinsma},\ and\ \citenamefont {Mason}}]{Zhao2012}%
  \BibitemOpen
  \bibfield  {author} {\bibinfo {author} {\bibfnamefont {K.}~\bibnamefont
  {Zhao}}, \bibinfo {author} {\bibfnamefont {R.}~\bibnamefont {Bruinsma}}, \
  and\ \bibinfo {author} {\bibfnamefont {T.~G.}\ \bibnamefont {Mason}},\
  }\href@noop {} {\bibfield  {journal} {\bibinfo  {journal} {Nat. Commun.}\
  }\textbf {\bibinfo {volume} {3}},\ \bibinfo {pages} {801} (\bibinfo {year}
  {2012})}\BibitemShut {NoStop}%
\bibitem [{\citenamefont {Bautista-Carbajal}\ and\ \citenamefont
  {Odriozola}(2014)}]{Bautista14}%
  \BibitemOpen
  \bibfield  {author} {\bibinfo {author} {\bibfnamefont {G.}~\bibnamefont
  {Bautista-Carbajal}}\ and\ \bibinfo {author} {\bibfnamefont {G.}~\bibnamefont
  {Odriozola}},\ }\href@noop {} {\bibfield  {journal} {\bibinfo  {journal} {J.
  Chem. Phys.}\ }\textbf {\bibinfo {volume} {140}},\ \bibinfo {pages} {204502}
  (\bibinfo {year} {2014})}\BibitemShut {NoStop}%
\bibitem [{\citenamefont {Bates}\ and\ \citenamefont
  {Frenkel}(2000)}]{Bates2000}%
  \BibitemOpen
  \bibfield  {author} {\bibinfo {author} {\bibfnamefont {M.~A.}\ \bibnamefont
  {Bates}}\ and\ \bibinfo {author} {\bibfnamefont {D.}~\bibnamefont
  {Frenkel}},\ }\href@noop {} {\bibfield  {journal} {\bibinfo  {journal} {J.
  Chem. Phys.}\ }\textbf {\bibinfo {volume} {112}},\ \bibinfo {pages} {10034}
  (\bibinfo {year} {2000})}\BibitemShut {NoStop}%
\end{thebibliography}

%

\end{document}